\documentclass[lettersize,journal]{IEEEtran}
\usepackage{amsmath,amsfonts}
\usepackage{algorithmic}
\usepackage{algorithm}
\usepackage{array}
\usepackage[caption=false,font=normalsize,labelfont=sf,textfont=sf]{subfig}
\usepackage{textcomp}
\usepackage{stfloats}
\usepackage{url}
\usepackage{verbatim}
\usepackage{graphicx}
\usepackage{cite}
\usepackage{color}
\usepackage{ulem}
\usepackage{bm}
\usepackage{booktabs}
\usepackage{diagbox}

\usepackage[mathscr]{eucal}
\usepackage{multirow}
\DeclareSubrefFormat{myparens}{#1(#2)}
\begin{document}

% \title{Deep Joint CSI Estimation-Feedback-Precoding for MU-MIMO OFDM Systems}
\title{Joint CSI Estimation-Feedback-Precoding via DJSCC for MU-MIMO OFDM Systems}

\author{Yiran Guo, Wei Chen,~\IEEEmembership{Senior Member,~IEEE}, Bo Ai,~\IEEEmembership{Fellow,~IEEE}, Lun Li

\thanks{
Yiran Guo, Wei Chen and Bo Ai are with the School of Electronic and Information Engineering, Beijing Jiaotong University, China. (Corresponding author: Wei Chen.)
Lun Li is with the State Key Laboratory of Mobile Network and Mobile Multimedia, Shenzhen, China, and ZTE Corporation, Nanshan District, Shenzhen 518055, China. 

This work is supported by ZTE Industry-University-Institute Cooperation Funds under Grant No. IA20240420002.
}
}


\maketitle

\begin{abstract}
As the number of antennas in frequency-division duplex (FDD) multiple-input multiple-output (MIMO) systems increases, acquiring channel state information (CSI) becomes increasingly challenging due to limited spectral resources and feedback overhead. 
In this paper, we investigate the impact of the feedback channel on CSI feedback in a multi-user MIMO orthogonal frequency-division multiplexing (OFDM) scenario, where the received downlink pilot signal is directly utilized as the source for CSI feedback in a joint design with CSI feedback and precoding. Considering the influence of the feedback channel, we propose an end-to-end joint CSI estimation–feedback–precoding network based on a deep joint source–channel coding architecture with an adaptive number of users.
% In this paper, we propose an end-to-end network that conducts joint design with pilot design, CSI estimation, CSI feedback, and precoding design in the multi-user MIMO orthogonal frequency-division multiplexing (OFDM) scenario. Multiple communication modules are jointly designed and trained with a common optimization objective to prevent mismatches between modules and discrepancies between individual module objectives and the final system goal. 
Experimental results demonstrate that, under the same feedback and CSI estimation overheads, the proposed joint multi-module end-to-end network achieves a higher multi-user downlink spectral efficiency than traditional algorithms based on separate architecture and partially separated artificial intelligence-based network architectures under comparable channel quality. Furthermore, compared to conventional separate architecture, the proposed network architecture with joint architecture reduces the computational burden and model storage overhead at the UE side, facilitating the deployment of low-overhead multi-module joint architectures in practice. Meanwhile, the network designed at the BS achieves user-number adaptability without increasing the number of trainable parameters, thereby reducing both model storage and distribution overhead by requiring only a single set of parameters for different numbers of users.
While slightly increasing storage requirements at the base station, it reduces computational complexity and precoding design delay, effectively reducing the effects of channel aging challenges.
\end{abstract}


\section{Introduction}
With the significant performance improvements brought by massive multiple-input multiple-output (MIMO) technology in 5G systems, and the growing demands for higher spectrum utilization, increased communication throughput, and faster transmission rates in the vision of 6G \cite{Survey_MIMO}, MIMO technology and its evolving variants are expected to become key technologies for 6G. The development of MIMO technology, such as extremely large-scale MIMO \cite{XL_MIMO} and reconfigurable intelligent surface \cite{RIS} technologies, offers valuable insights into the application of MIMO in 6G systems. Additionally, research into terahertz band communication \cite{THz} provides theoretical support for MIMO systems that can accommodate more antennas and more effectively exploit spatial resources. However, to fully harness the potential of MIMO technology, obtaining channel state information (CSI) with low overhead is a critical challenge. In time-division duplex (TDD) systems, channel reciprocity between uplink and downlink channels operating in the same frequency band simplifies downlink CSI acquisition. Conversely, in frequency-division duplex (FDD) systems, where channel reciprocity is weak, the increased number of antennas presents significant challenges for downlink CSI estimation (CE) and feedback with low overhead and low latency.

Traditional methods for CSI feedback include the codebook-based feedback method, commonly applied in 5G, and the compressed sensing (CS) method based on sparsity assumptions \cite{Sparse_Assume}. However, the codebook-based feedback approach faces scalability challenges as the number of antennas increases. Ensuring the same reconstruction accuracy requires a larger codebook size, leading to increased complexity in codeword search and higher feedback overhead. Similarly, the CS-based CSI feedback method is constrained by the sparsity assumption, making it difficult to achieve theoretical performance. Additionally, the complex reconstruction algorithms in CS methods introduce significant computational overhead and delay, making them unsuitable for fast time-varying channels prone to CSI aging.

With the promising performance of artificial intelligence (AI) techniques in fields such as image compression and transmission \cite{AI_Image_Process}, and signal processing \cite{AI_Signal_Process}, AI-based methods are increasingly adopted for communication tasks, particularly CSI feedback \cite{CsiNet, CSI_FB_Overview, One_Bit_Superimposed_CSI_FB, LOS_Superimposed_CSI_FB, GAT_Superimposed_CSI_FB}. 
In \cite{One_Bit_Superimposed_CSI_FB}, the feedback overhead can be reduced by superimposing downlink CSI onto the uplink signal; however, this approach introduces a new challenge, i.e., interference between the superimposed CSI feedback and the original uplink signal. To mitigate this challenge, \cite{One_Bit_Superimposed_CSI_FB} proposed an AI-based fusion learning scheme using 1-bit CS, which leverages the partial reciprocity of uplink and downlink channels for CSI feedback and reconstruction. Another line of research retains the resources allocated for CSI feedback but seeks to enable the extraction of as many low-dimensional CSI features as possible without compromising reconstruction performance through advanced network design or the incorporation of auxiliary information, balancing low overhead and high reconstruction accuracy.
% Another line of research retains the resources allocated for CSI feedback but aims to balance low overhead with high reconstruction accuracy by enabling the extraction of low-dimensional CSI features through advanced network design or the incorporation of auxiliary information.
These methods treat the CSI as an image, where the real and imaginary components are considered as separate channels. AI-based techniques exploit the correlation of CSI in the frequency and spatial domains to compress high-dimensional CSI into low-dimensional feature vectors using an auto-encoder at the transmitter. The compressed features are fed back, and the CSI is reconstructed at the receiver using a corresponding decoder. This CSI feedback process is considered an image compression and reconstruction task, trained end-to-end from user equipment (UE) to the base station (BS), with reconstruction accuracy as the optimization objective to update trainable parameters. By replacing traditional source coding and decoding modules with AI-based compression and reconstruction modules, various network architectures have been proposed to achieve low-overhead, high-accuracy CSI feedback, addressing key limitations of traditional methods \cite{CsiNet_plus, SwinTransformer, Transformer_CSI}.

The use of AI networks to replace traditional source coding and decoding modules in communication architectures is referred to as a separate source-channel coding (SSCC) architecture. In this approach, the design of the separate architecture allows the network to be trained without considering the impact of channel coding and decoding on performance, which results in the network performance being highly affected by channel quality.
% As a result, when the network is tested with actual channel effects, if the channel quality is insufficient and beyond the capacity of the channel coding, the inputs to the source decoding module will contain a significant number of errors. This discrepancy causes the distribution of the inputs to differ from the distribution seen during training, leading to severe reconstruction errors. This phenomenon, often called the ``cliff effect'', occurs when the system’s performance sharply degrades due to mismatched input distributions. To address this issue, 
In contrast, the authors of \cite{ADJSCC} proposed a deep joint source-channel coding (DJSCC) architecture for CSI feedback, which jointly considers source coding, channel coding, channel decoding, and source decoding. This approach abandons the traditionally modular design seen in communication systems. By leveraging the powerful nonlinear capabilities of AI, the DJSCC module provides more stable and reliable CSI reconstruction performance compared to the SSCC architecture.

\begin{table*}[!t]
\renewcommand{\arraystretch}{1.05}
\centering
\caption{Comparison of Existing Studies and the Proposed Network Architecture}
\label{Architect_Comparison}
\begin{tabular}{c|c|c|c|c|c|c|c|c}
\toprule
\textbf{Ref.} &
  \textbf{\begin{tabular}[c]{@{}c@{}}Pilot\\ Design\end{tabular}} &
  \textbf{\begin{tabular}[c]{@{}c@{}}Channel\\ Estimation\end{tabular}} &
  \textbf{\begin{tabular}[c]{@{}c@{}}CSI\\ Compression\end{tabular}} &
  \textbf{\begin{tabular}[c]{@{}c@{}}Channel\\ Coding\end{tabular}} &
  \textbf{\begin{tabular}[c]{@{}c@{}}Channel\\ Decoding\end{tabular}} &
  \textbf{\begin{tabular}[c]{@{}c@{}}CSI\\ Construction\end{tabular}} &
  \textbf{\begin{tabular}[c]{@{}c@{}}Precoding\\ Design\end{tabular}} &
  \textbf{\begin{tabular}[c]{@{}c@{}}Feedback\\ Method\end{tabular}} 
   \\ \midrule
\cite{CsiNet_plus} & & & \checkmark & & & \checkmark & & Quantification +   Ideal Feedback 
   \\ 
\cite{ADJSCC} & & & \checkmark & \checkmark & \checkmark & \checkmark & & DJSCC 
   \\ 
\cite{JFPNet} & & & \checkmark & \checkmark & \checkmark & \checkmark & \checkmark & DJSCC 
   \\ 
\cite{Joint_Pilot_CE_FB_BF} & & \checkmark & \checkmark & & & & \checkmark & Ideal   Feedback 
   \\ 
\cite{CEFNet} & & \checkmark & \checkmark & & & \checkmark & \checkmark & Quantification +   Ideal Feedback 
   \\ 
\cite{ICC_JEFPNet} & \checkmark & \checkmark & \checkmark & & & \checkmark &  \checkmark & Quantification +   Ideal Feedback
   \\ 
\cite{YuWei_JEFPNet} & \checkmark & \checkmark & \checkmark & & & \checkmark & \checkmark & Quantification +   Ideal Feedback 
   \\ 
\cite{Joint_PD_FB_BF} & \checkmark & \checkmark & \checkmark & & & \checkmark & \checkmark & Quantification +   Ideal Feedback 
   \\ 
\cite{MU_Adaptive} & \checkmark & \checkmark &  \checkmark &   &   &  \checkmark &  \checkmark &  Quantification +   Ideal Feedback 
   \\
   \cmidrule{1-9}
   \textbf{Proposed} &  \checkmark &  \checkmark &  \checkmark &   \checkmark &   \checkmark &  \checkmark &  \checkmark &  DJSCC
  \\
   \toprule
\end{tabular}
\end{table*}
The challenges encountered by separate architectures also arise in both the upstream task (CE) and the downstream task (precoding design) of CSI feedback. Inaccurate CE can cause a mismatch in the input distribution of the CSI compression module between the training and testing phases, leading to inaccurate CSI feedback. 
% To address this issue, \cite{CEFNet} proposed a joint CE (CE) and CSI feedback architecture. Similarly, inaccurate CSI feedback can lead to the following precoding design task, thereby reducing downlink channel capacity.  In \cite{ICC_JEFPNet, YuWei_JEFPNet}, the authors consider the CSI feedback, CE, and precoding design tasks simultaneously, but overlook the impact of channel coding and the actual channel conditions on performance.  
In \cite{CsiNet_plus, Implicit_CSI_FB}, AI-based two-sided CSI compression and reconstruction modules are designed and jointly trained to perform the CSI feedback task, typically resulting in improved CSI reconstruction performance compared to one-sided feedback architectures in \cite{CsiNet, CFnet} and codebook-based feedback architectures in traditional algorithms. In \cite{ADJSCC}, the CSI compression and reconstruction module is jointly designed with the channel coding and decoding module to mitigate the ``cliff effect'' and reduce retransmission delays. In our previous work \cite{JFPNet}, we proposed a joint feedback and precoding network (JFPNet) for multi-user (MU) MIMO scenarios that simultaneously addresses both CSI feedback and its downstream task, precoding design. The source coding and channel coding modules are jointly designed on the UE side, while the channel decoding, source decoding, and precoding design modules for MU scenarios are jointly designed on the BS side. According to the experimental results in \cite{JFPNet}, this joint multi-module architecture outperforms the separate architecture for the CSI feedback and precoding design task. However, this work assumes that acquiring downlink CSI at the UE is ideal, and that multiple networks must be trained and stored to accommodate different numbers of users. In \cite{Joint_Pilot_CE_FB_BF}, an end-to-end multi-module joint architecture was designed to simultaneously address upstream and downstream tasks, including pilot length design, CE, CSI feedback, and beamforming precoding. However, its pilot sequences are randomly generated from a complex unit sphere, limiting the ability to fully utilize channel information and exploit the learning capacity of the end-to-end architecture. Additionally, its downlink CE process relies on artificially introduced noise to simulate CE errors, which are inconsistent with the actual errors caused by specific CE methods, resulting in performance degradation. Both the joint CSI feedback and estimation architecture in \cite{CEFNet} and the joint CE, feedback, and precoding design architecture in \cite{ICC_JEFPNet, YuWei_JEFPNet, Joint_PD_FB_BF, MU_Adaptive} address the end-to-end performance loss caused by mismatches between the distribution of the estimated and trained CSI. However, both overlook the potential ``cliff effect'' caused by the omission of channel coding and decoding
, as shown in Table.~\ref{Architect_Comparison},
as well as the added retransmission overhead and delay when channel quality degrades.

The impact of the ``cliff effect'' in CSI feedback on system performance has been thoroughly evaluated in~\cite{ADJSCC}, and the proposed DJSCC architecture offers a new solution paradigm to mitigate this problem \cite{JFPNet,DJSCC_FB_1, DJSCC_FB_2}. However, the system architecture remains to be explored when the source changes from the ideally acquired CSI, as considered in~\cite{ADJSCC,JFPNet,DJSCC_FB_1}, to the more realistic non-ideal CSI obtained through CE or directly from the received pilot signal. Particularly in communication scenarios with a variable number of users, designing a bilateral framework for CE, CSI feedback, and precoding that is both user-number-adaptive and deployment-friendly is crucial to advancing the practical application of the DJSCC architecture. To address this challenge, and with channel coding taken into account, we propose a joint design of CE, CSI feedback, and precoding with user-number-adaptive capabilities. In summary, our contributions to this work are as follows:
\begin{itemize}
\item[$\bullet$]In a MU MIMO orthogonal frequency-division multiplexing (OFDM) system, under consideration of the impact of channel coding on network performance, we propose an DJSCC-based end-to-end network that jointly addresses CSI feedback, its upstream task (CE), and its downstream task (precoding design). The proposed framework considers the entire closed-loop system, starting from the pilot design at the BS, to the CE and joint source-channel coding at the UE, and finally to the joint source-channel decoding and precoding design at the BS. Only one joint multi-module designed network is deployed at each device terminal.

\item[$\bullet$]Considering a variable number of users, we design a joint CE, CSI feedback, and precoding network based on a Transformer architecture \cite{Transformer} that is adaptive to the number of UE. The number of stored parameters in the proposed network remains constant regardless of the number of users, and within a defined user range, its performance approximates that of networks trained with a fixed number of users, eliminating the need to switch models. This design significantly reduces model storage overhead, switching delays and model distribution overhead form the BS to the UE when the number of users changes.

\item[$\bullet$]Performance comparisons between joint and separate architectures with DJSCC architecture under different numbers of UEs are conducted through simulation experiments. By jointly designing the pilot, CE, and CSI compression modules, the proposed framework extracts task-oriented semantic information for transmission via end-to-end training, thereby enhancing feedback efficiency. The simulation results demonstrate both the performance gains and the stability of the joint architecture. In particular, the proposed JEFPNet exhibits stronger adaptability than separate architectures, especially in the presence of significant CE and feedback errors, and is capable of designing more efficient precoding vectors. Moreover, by avoiding explicit CE at the UE, JEFPNet reduces both storage, distribution, and computational overhead.

\end{itemize}

The remainder of this paper is organized as follows: Section \ref{section1} introduces the system model for FDD MU-MIMO-OFDM systems and provides research background on AI-based CSI feedback to support this work. Section \ref{section2} describes the proposed framework, including the network architectures and end-to-end optimization objectives. Section \ref{section3} presents the experimental results, and Section \ref{section4} concludes the paper.

\textit{Notations:}In this paper, we denote Vectors and Matrices with boldface lower- and upper-case letters, respectively. ${\mathbf{A}}^T$ and ${\mathbf{A}}^H$ are transpose of ${\mathbf{A}}$ and complex conjugate transpose operation of ${\mathbf{A}}$, respectively. $a^*$ denotes the conjugate processing of complex variable $a$. ${\left\| \cdot \right\|_1}$, ${\left\| \cdot \right\|_2}$ and $ \lVert \cdot  \rVert $ represents the $\ell\text{-}1$ norm, the $\ell\text{-}2$ norm and modulus operations on complex numbers, respectively. $\mathbb{E}\left(\cdot \right)$ denotes the statistical expectation. $\{\cdot, \cdot\}$ denotes a set containing multiple elements. $\{\mathbf{A}_k\}_{k=1}^K$ denotes a set containing $K$ elements from $\mathbf{A}_1$ to $\mathbf{A}_K$, and $\{\mathbf{A}(k)\}_{k=1}^K$ denotes a set containing $K$ elements from $\mathbf{A}(1)$ to $\mathbf{A}(K)$.The symbols $\mathbb{R}$ and $\mathbb{C}$ represent sets of real and complex numbers, respectively. $\lfloor{A}\rfloor$ indicates the floor operation.
% $\odot$ is hadamard product. $diag\left({\cdot}\right)$ denotes the diagonal matrix formed by the numbers in parentheses, and the numbers in parentheses are the diagonal elements.

\section{System Model and Background}
\label{section1}
%%%%%%%%%%%%%%%%%%%%%%%%%%%%%%%%%
\subsection{System Model}
Given that OFDM systems underpin most contemporary broadband wireless standards (e.g., 
% \textcolor{blue}{long term evolution (LTE), 5G new radio (NR)}
long term evolution (LTE), 5G new radio (NR)
)   and offer well-established advantages in multiuser MIMO scenarios, we adopt a MIMO--OFDM system with $K$ users and a maximum of $K_{\max}$ users for our proposed scheme to ensure practical deployability. Each user is equipped with a single omnidirectional antenna, while the BS employs an $N_t$-element antenna array. Although one could first investigate a narrowband system and subsequently extend it to a wideband OFDM system through a parallel scheme, such an approach fails to exploit frequency-domain correlation and introduces additional learning parameters. Therefore, we directly study the problem of multiuser precoding in a wideband system. The system operates with $N_c$ subcarriers. In this system, we simultaneously consider downlink CE, downlink CSI feedback in the uplink channel, and downlink precoding design. The downlink CSI of the $k$-th user can be expressed as $\mathbf{H}_{d}(k) \in \mathbb{C}^{N_c \times N_t}$, and the corresponding uplink feedback channel can be expressed as $\mathbf{H}_{u}(k) \in \mathbb{C}^{N_c \times N_t}$. 

For the downlink CE phase, we consider the pilot-assisted CE method, that downlink CSI is estimated at the UE with the help of the known pilot sequence for transmitter and receiver. For the pilot sequence, in the time domain, we use sequential $L$ OFDM symbols for CE. To reduce pilot overhead, pilots are placed at equal $g$ intervals in the frequency domain, while the non-pilot positions are utilized for data or other channel control information transmission. $M$ subcarriers are sampled at equal intervals $g$ from the $N_c$ subcarriers to place the pilots, and the indices of these subcarriers form the set of pilot sequences, denoted as $ \mathcal{M} = \{\mathcal{M}_1, \cdots, \mathcal{M}_M\} \in \mathbb{R}$, where $M = \lfloor{N_c / g}\rfloor$. In this paper, user interference during feedback is not considered. Consequently, the same pilot symbols are employed across different users. The pilot sequence at the $\mathcal{M}_m$-th subcarrier can be expressed as $\mathbf{P}_{\mathcal{M}_m} =\left[{\mathbf{p}_{\mathcal{M}_m}^{(1)}, \cdots, \mathbf{p}_{\mathcal{M}_m}^{(L)}}\right] \in \mathbb{C}^{N_t \times L}$.

The received pilot signals at the $\mathcal{M}_m$-th subcarrier of the $k$-th user can be expressed as
\begin{equation}
	\label{Received_pilot}
	\mathbf{y}_{\mathcal{M}_m}(k) = \mathbf{h}^T_{d,{\mathcal{M}_m}}(k)\mathbf{P}_{\mathcal{M}_m}+\mathbf{n}_{d,\mathcal{M}_m}(k) \in \mathbb{C}^{1 \times L},
\end{equation}
where $\mathbf{h}_{d,{\mathcal{M}_m}} \in \mathbb{C}^{N_t}$ denotes the ${\mathcal{M}_m}$-th subcarrier of downlink channel $\mathbf{H}_{d}=\left[{\mathbf{h}_{d,1},\cdots, \mathbf{h}_{d,N_c}}\right]^T$. $\mathbf{n}_{d,\mathcal{M}_m}(k)$ is downlink additive white Gaussian noise (AWGN) with mean zero and variance $\sigma_{ce}^{2}$. In summary, the downlink pilot transmission process can be expressed as follows:
\begin{equation}
	\label{DL_P_Trans}
	\mathbf{Y}(k)= \mathcal{F}_{\mathrm{ce},k}\left({\mathscr{P}}\right),
\end{equation}
where $\mathbf{Y}(k)=\left[{\mathbf{y}_{\mathcal{M}_1}^T(k), \cdots, \mathbf{y}_{\mathcal{M}_M}^T(k)}\right]^T \in \mathbb{C}^{M \times L}$ represents the received pilot signal, and $\mathscr{P}=\{\mathbf{P}_{\mathcal{M}_m}\}_{m=1}^{M}$ presents all pilot symbols on the selected subcarrier and OFDM symbols. $\mathcal{F}_{\mathrm{ce},k}\left(\cdot\right)$ denotes the downlink pilot signal transmission process of the $k$-th user.
Notably, the pilot sequence of each OFDM symbol satisfies the power constraint
\begin{equation}
	\label{pilot_constrain}
	\frac{1}{M}\sum\limits_{\mathcal {M}_m \in \mathcal{M}} {\left\| {{\bf{p}}_{{{\mathcal {M}}_m}}^{(l)}} \right\|_2^2}  = 1.
\end{equation}
With the channel normalization, the downlink signal-to-noise rate (SNR) during the CE phase can be expressed as
\begin{equation}
	\label{CE_SNR}
	\mathrm{SNR_{ce}}=\frac{1}{\sigma_{ce}^{2}}.
\end{equation}

Using the received pilot signal $\mathbf{Y}(k)$ and the known pilot sequences $\mathscr{P}$, the downlink CSI of the $k$-th user can be estimated and subsequently compressed to obtain a semantic representation of the downlink CSI for reconstruction, denoted as
\begin{equation}
	\label{CE_Compression}
	\mathbf{s}(k) = \mathcal{E}_{\theta}\Big({\mathbf{Y}(k), \mathscr{P}}\Big) \in \mathbb{C}^{Z},
\end{equation}
where $\mathcal{E}_{\theta}\left({\cdot}\right)$ denotes the processing at the UE side, including downlink CSI CE and CSI semantic information extraction, and $\theta$ denotes the set of trainable parameters for this process. $Z$ denotes the number of feature elements, which also corresponds to the number of subcarriers occupied by the feedback after OFDM mapping. It is worth emphasizing that the output features satisfy the power constraints, as
\begin{equation}
	\label{FB_constrain}
	\frac{1}{Z}{\left\| {\mathbf{s}(k)} \right\|_2^2}  = 1.
\end{equation}

The received CSI semantic feature vector of the $z$-th subcarrier can be expressed as
\begin{equation}
	\label{Received_CSI}
	\mathbf{y}_{z}(k) = \mathbf{h}_{u,z}^{T}(k)s_z(k)+\mathbf{n}_{u,z}(k) \in \mathbb{C}^{N_t},
\end{equation}
where $s_z(k)$ is the $z$-th element of $\mathbf{s}(k)$, and $\mathbf{h}_{u,z}(k) \in \mathbb{C}^{N_t}$ is $z$-th row of uplink channel $\mathbf{H}_{u}(k)$ for the $k$-th user. $\mathbf{n}_{u,z}(k)$ is uplink AWGN during the CSI feedback phase with zero mean and $\sigma_{u}^{2}$ variance. With the uplink noise power and transmission power constrain in Eq.~\eqref{FB_constrain}, uplink feedback SNR can be expressed as
\begin{equation}
	\label{FB_SNR}
	\mathrm{SNR_{u}}=\frac{1}{\sigma_{u}^{2}}.
\end{equation}

At the BS, we use a maximal ratio combining (MRC) receiver for detecting signals. The combining matrix is derived by the estimated uplink CSI. In this paper, we assume perfect uplink CE. Therefore, the combining matrix of the $z$-th subcarrier $\mathbf{W}_z(k)$ can be expressed as
\begin{equation}
	\label{combining_matrix}
	\mathbf{W}_z(k) = \frac{\mathbf{h}_{u,z}(k)}{\left\|{\mathbf{h}_{u,z}(k)}\right\|_2}.
\end{equation}

The detected semantic information $\hat{\mathbf{s}}(k)$ of downlink CSI can be expressed as
\begin{equation}
	\label{detected_CSI}
	\hat{s}_z(k)=\mathbf{W}_z^H(k)\mathbf{y}_{z}(k),
\end{equation}
where $\hat{s}_z(k)$ represents the $z$-th element of the vector $\hat{\mathbf{s}}(k)$. The uplink CSI feedback process can be obtained with
\begin{equation}
	\label{UL_CSI_FB}
	\hat{\mathbf{s}}(k)=\mathcal{F}_{\mathrm{u},k}\left({\mathbf{s}(k)}\right),
\end{equation}
where $\mathcal{F}_{\mathrm{u},k}\left(\cdot\right)$ denotes uplink data transmission function of the $k$-th user.

The semantic feature information of downlink CSI from MU is aggregated at the BS. This aggregated downlink channel semantic feature information, encompassed from various users, is subsequently used to enable a joint precoding design in the MU scenarios. The operation at the BS can be expressed as follows:
\begin{equation}
	\label{Joint_CSI_recon_Pre}
	\mathscr{V} = \mathcal{D}_{\psi}\Big({\{{\hat {\mathbf{s}}(k)}\}_{k=1}^{K}}\Big),
\end{equation}
where $\mathscr{V}=\{{\mathbf{V}(k)}\}_{k=1}^K$  represent the set of precoding matrices designed for different users, and $\mathbf{V}(k)=\left[{\mathbf{v}_{1}(k), \cdots, \mathbf{v}_{N_c}(k)}\right]^T \in \mathbb{C}^{N_c \times N_t}$ denotes precoding matrix of the $k$-th user. $\mathcal{D}_{\psi}(\cdot)$ presents the process of downlink CSI reconstruction and precoding design with the trainable parameter set $\psi$ at the BS side. It should be emphasized that the precoding matrices for different users satisfy the power constraint, as
\begin{equation}
	\label{Precoding_constraint}
	\sum\limits_{k=1}^{K}{\left\|{\mathbf{v}_{n_c}(k)}\right\|_{2}^{2}}=P,
\end{equation}
where the subscript $n_c \in \left\{{1,\cdots,N_c}\right\}$ represents the $n_c$-th subcarriers, and $P$ denotes downlink transmission power. The downlink transmit signal $x_{n_c}(k)$ of the $k$-th user at the $n_c$-th subcarrier after precoding process can be represented as $\mathbf{v}_{n_c}(k)x_{n_c}(k)$. With the transmit signals satisfying statistically independent and $\mathbb{E}\left\{{x_{n_c}(k)x^*_{n_c}(k)}\right\}=1$, the SNR of the downlink channel after precoding process can be expressed as
\begin{equation}
	\label{Pre_SNR}
	\mathrm{SNR_{d}}=\frac{P}{\sigma_{d}^{2}}.
\end{equation}

Correspondingly, the received signal of the $k$-th user is
\begin{equation}
	\label{Precoding_receive}
	r_{n_c}(k)=\mathbf{h}^T_{d,n_c}(k)\sum\limits_{k=1}^K{\mathbf{v}_{n_c}(k)x_{n_c}(k)}+n_{n_c}(k),
\end{equation}
where $n_{n_c}(k)$ represents the downlink AWGN with power $\sigma_{d}^2$ during the data transmission phase. The downlink spectral efficiency is given in Eq.~\eqref{Sum_rate}.

It is worth emphasizing that, for simplicity, we considers a fully digital beamforming architecture in this paper, where the number of radio frequency (RF) chains $N_{\mathrm{RF}}$ is equal to the number of BS antennas $N_t$. If the proposed network is to be applied in a hybrid beamforming architecture, the analog precoder $\mathbf{F}_{\mathrm{RF}} \in \mathbb{C}^{N_t \times N_{\mathrm{RF}}}$ and the digital precoder $\mathbf{F}_{\mathrm{BB}} \in \mathbb{C}^{N_{\mathrm{RF}} \times K}$ should be designed separately for effective deployment.
\begin{figure*}[ht]
\centering
\begin{equation}
	\label{Sum_rate}
	R_{\theta, \psi,\mathscr{P}} = \sum\limits_{k=1}^{K}{\sum\limits_{n_c=1}^{N_c}{\mathrm{log}_2\left({1+\frac{\lVert{\mathbf{h}_{d, n_c}^T(k)\mathbf{v}_{n_c}(k)x_{n_c}(k)}\rVert^2}{\sum\limits_{\substack{m \neq k}}{\lVert{\mathbf{h}_{d, n_c}^T(k)\mathbf{v}_{n_c}(m)x_{n_c}(m)}\rVert^2}+\sigma_d^2}}\right)}}.
\end{equation}
\hrulefill
\vspace*{8pt}
\end{figure*}

\subsection{Background Description}
\begin{figure}[!t]
  \centering
  \subfloat[]{
    \label{SSCC}
    \includegraphics[width=3.3in]{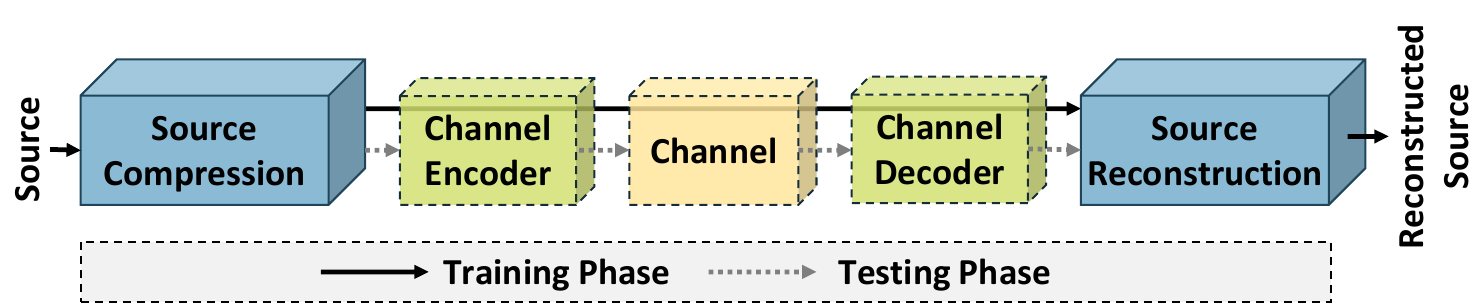}
    
  }\\
  \subfloat[]{
    \label{DJSCC}
    \includegraphics[width=3.3in]{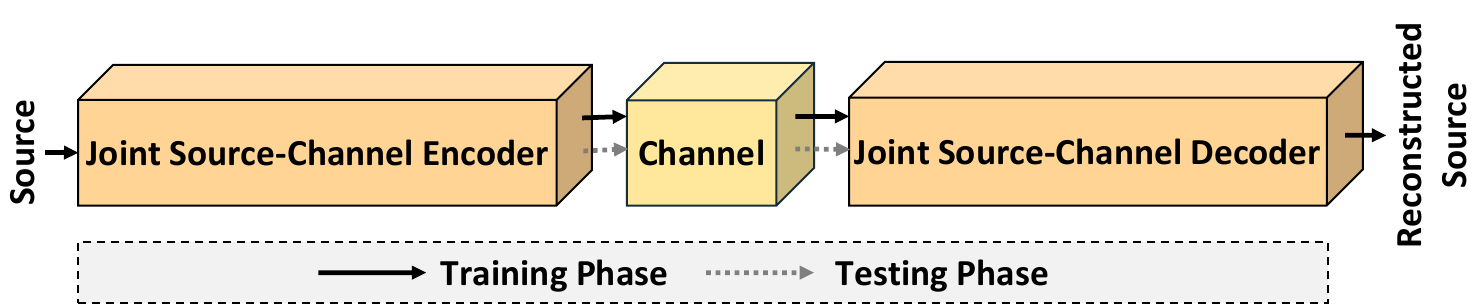}
    
  }\\
  \caption{Architectures of AI-based CSI feedback: \protect\subref{SSCC} the separate architecture; \protect\subref{DJSCC} the joint architecture.}
  \label{construction}
\end{figure}

The AI-based CSI feedback architecture can be classified into separate and joint architectures, as illustrated in Fig.~\ref{construction}. In the separate architecture, the AI network is designed to focus exclusively on source compression and reconstruction tasks, employing an end-to-end training approach that neglects the impact of channel coding and decoding on overall performance. As a consequence, when channel coding, decoding, and the communication channel are taken into account during testing, a phenomenon known as the ``cliff effect'' occurs. This effect arises when the channel quality deteriorates beyond the capacity of the channel coding, leading to a sharp decline in source reconstruction quality. This phenomenon will result in system performance degradation or retransmission delay. The root cause is the alteration of the input distribution to the source reconstruction module, as the inclusion of channel coding, decoding, and transmission changes the input characteristics. During training in the separate architecture, the output from the source compression module is directly fed into the source reconstruction module, maintaining consistency in the distributions between the outputs of the CSI compression module and the inputs of the CSI reconstruction module. However, during testing, if channel coding fails, significant errors are introduced into the inputs of the source reconstruction module, causing a mismatch between the input distribution observed during training and that encountered during testing, resulting in performance degradation.

In contrast, the joint architecture integrates source and channel coding into a unified module, allowing the network to adapt to changing channel conditions during training via an end-to-end approach. This integration effectively mitigates the ``cliff effect'' typically observed in separate architectures. The joint architecture has been successfully implemented in several of our previous works for the CSI feedback task\cite{ADJSCC, JFPNet} and the image transmission task \cite{AI_Image_Process}, demonstrating its efficacy. Consequently, the feedback mechanism in this paper also adopts this DJSCC framework.

\section{DJSCC-Based MU Joint CE, Feedback, and Precoding Design}
\label{section2}
\begin{figure*}[!t]
	\centering
	\includegraphics[width=6.0in]{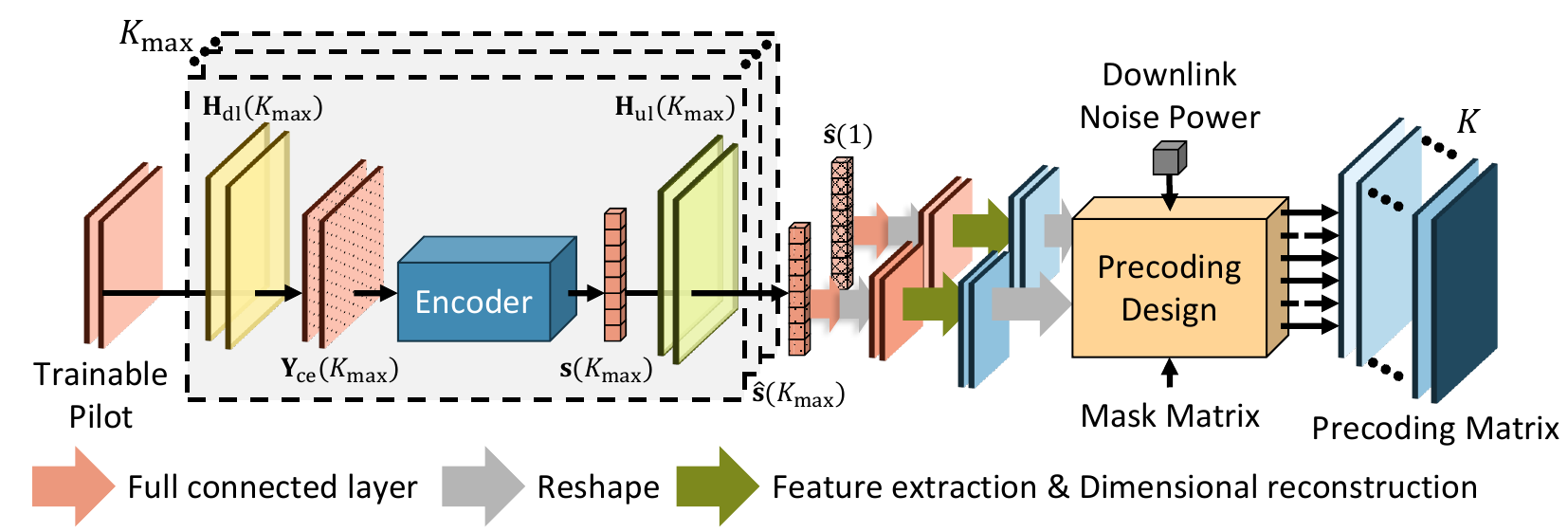}
	\caption{Architecture of the JEFPNet. (Certain parameter-free operations are omitted for simplicity.)}
	\label{E2E_arct_JEFPNet}
\end{figure*}

In the MU MIMO-OFDM scenario, we propose an end-to-end joint estimation, feedback and precoding network (JEFPNet), as shown in Fig.~\ref{E2E_arct_JEFPNet}, by considering the three processes of downlink CE, uplink CSI feedback under the DJSCC architecture, and downlink precoding design. To ensure that the proposed network adapts to varying numbers of users, a random user activation mask matrix is introduced in the precoding stage.

The downlink CE adopts the pilot-assisted CE algorithm, and the pilot symbols use trainable pilot symbols. A complex fully connected (FC) neural network architecture is employed to facilitate pilot design. Specifically, the complex-valued pilot parameters to be trained are separated into real and imaginary components. The real and imaginary parts of the pilot sequence for the $\mathcal{M}_m$-th subcarrier can then be expressed as $\mathcal{R}\left({\mathbf{P}_{\mathcal{M}_m}}\right)$ and $\mathcal{I}\left({\mathbf{P}_{\mathcal{M}_m}}\right)$. Similarly, the downlink channel and the received signal can be expressed in the same form, the real part of the received signal of the $k$-th user in the $\mathcal{M}_m$-th subcarrier can be expressed as
\begin{equation}
	\label{R_Recive_real}
 \begin{split}
	\mathcal{R}\big({\mathbf{y}_{\mathcal{M}_m}(k)}\big)=& \mathcal{R}\big({\mathbf{h}^T_{d,\mathcal{M}_m}(k)}\big)\mathcal{R}\big({\mathbf{P}_{\mathcal{M}_m}}\big)-\\
 & \mathcal{I}\big({\mathbf{h}^T_{d,\mathcal{M}_m}(k)}\big)\mathcal{I}\big({\mathbf{P}_{\mathcal{M}_m}}\big)+\\&\mathcal{R}\big({\mathbf{n}_{d,\mathcal{M}_m}(k)}\big),
 \end{split}
\end{equation}
while the imaginary part can be expressed as
\begin{equation}
	\label{R_Recive_imag}
 \begin{split}
	\mathcal{I}\big({\mathbf{y}_{\mathcal{M}_m}(k)}\big)=& \mathcal{R}\big({\mathbf{h}^T_{d,\mathcal{M}_m}(k)}\big)\mathcal{I}\big({\mathbf{P}_{\mathcal{M}_m}}\big)+\\
 & \mathcal{I}\big({\mathbf{h}^T_{d,\mathcal{M}_m}(k)}\big)\mathcal{R}\big({\mathbf{P}_{\mathcal{M}_m}}\big)+\\&\mathcal{I}\big({\mathbf{n}_{d,\mathcal{M}_m}(k)}\big).
 \end{split}
\end{equation}

\begin{figure*}[ht]
\centering
\begin{equation}
\label{Optimization_Objective}
\begin{split}
	\mathop {{\rm{maximize}}}\limits_
        {\theta,\psi, \mathscr{P}} & R_{\theta, \psi,\mathscr{P}}= \sum\limits_{k=1}^{K_{\max}}{\sum\limits_{n_c=1}^{N_c}{{\Upsilon _k}\mathrm{log}_2\left({1+\frac{\lVert{\mathbf{h}_{d, n_c}^T(k)\mathbf{v}_{n_c}(k)x_{n_c}(k)}\rVert^2}{\sum\limits_{\substack{m \neq k}}{\lVert{\mathbf{h}_{d, n_c}^T(k)\mathbf{v}_{n_c}(m)x_{n_c}(m)}\rVert^2}+\sigma_d^2}}\right)}}, \\
        \text{s.t.} & \quad 
        \left\{{\left\{{\mathbf{v}_{n_c}(k)}\right\}}_{{n_c}=1}^{N_c}\right\}_{k=1}^K=\mathcal{D}_{\psi}\left({\left\{{\mathcal{F}_{\mathrm{u},k}{\left({\mathcal{E}_{\theta}{\left({\mathcal{F}_{\mathrm{ce},k}{\left({\mathscr{P}}\right)}}, \mathscr{P}\right)}}\right)}}\right\}_{k=1}^{K}}\right),
\end{split}
\end{equation}
\hrulefill
\vspace*{8pt}
\end{figure*}

\begin{figure}[!t]
	\centering
	\includegraphics[width=3.4in]{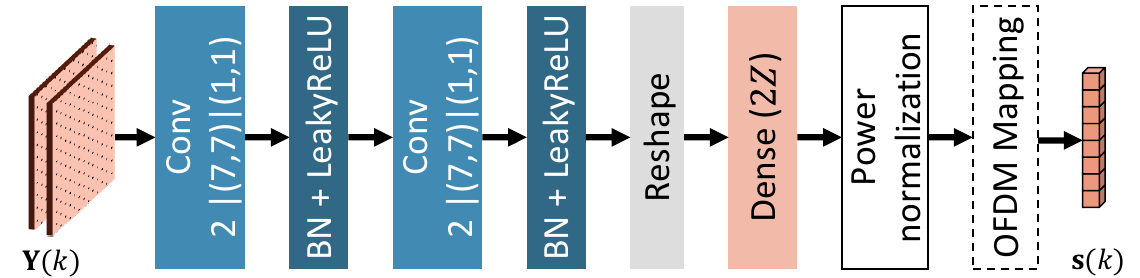}
	\caption{Architecture of encoder module. Different users utilize encoders with shared parameters, meaning the input tensor and output vector do not include user-specific identifiers.}
	\label{Encoder}
\end{figure}

The received pilot signal vectors are concatenated in the frequency domain to form the received pilot matrix, denoted as $\mathbf{Y}(k)$ for the $k$-th user. The real and imaginary parts of the received signal are treated as separate channels, and the received pilot signal---representing the task-oriented feature information and containing noise---is used as the source input to the DJSCC encoder for direct compression. The network architecture of the encoder module is shown in Fig.~\ref{Encoder}. Here, ``$\mathrm{Conv} \medspace F|(W_1,W_2)|(S_1, S_2)$'' indicates a convolutional layer with $F$ filters, a window size is $W_1 \times W_2$, and a stride of $S_1 \times S_2$. ``$\mathrm{BN}+\mathrm{LeakyReLU}$'' represents a batch normalization (BN) layer followed by the ``$\mathrm{LeakyReLU}$'' activation function. ``$\mathrm{Dense}(2Z)$'' denotes a FC layer with $2Z$ neurons. The detailed network parameters are specified in the Fig.~\ref{Encoder}, and the number of layers and the convolutional kernel size are consistent with those commonly adopted in existing CSI feedback studies \cite{CsiNet_plus, ADJSCC, JFPNet}. In the encoder module, CSI features are first extracted by the convolutional layer, then reshaped into vectors, and finally compressed into $2Z$-dimensional vectors using an FC layer with $2Z$ neurons, which corresponds to the feedback bandwidth $Z$ after real-complex transformation. The transmit power is fixed through power normalization and OFDM mapping of the feature vectors before transmission.

It is important to note that users are assumed to be located within the same cell, sharing similar scatterer environments and channel statistical parameters. Consequently, the downlink CSI of different users follows the same distribution, allowing us to use a single trainable pilot across users. Additionally, the encoder module parameters are also shared among other users, significantly reducing the number of training parameters and accelerating training convergence. This parameter-sharing strategy also eliminates the need for the BS to adjust the pilot sequences for different users. Additionally, the UE only needs to store the encoder parameters for a specified cell, or it can receive them directly from the BS without requiring a user identity document.

In this paper, we do not consider multiple access scenarios. We assume that each user feeds back CSI during a single OFDM symbol, with CSI feedback occurring sequentially for each user without inter-user interference during the feedback phase.
The received compressed signals for each user are first dimensionally recovered using a FC layer with $2MN_t$ neurons. The downlink CSI semantic feature tensor, with dimensions $M \times N_t \times 2$, is then reconstructed through a ``$\mathrm{Reshape}$'' layer. Therefore, the CSI for each user can initially be regarded as a semantic feature matrix of size $M \times N_t \times 2$ at the BS. The neuron parameters are configured with reference \cite{ADJSCC} to align with the separation architecture, thereby ensuring a fair comparison.
This MU CSI semantic feature tensor is processed by a convolutional network and a transposed convolutional network to extract features and reconstruct dimensions aligned with the dimensional requirements of precoding design. Finally, the resulting feature tensor, combined with downlink noise power and mask matrix, is fed into the precoding design module to complete the precoding design. The outputs of the precoding design module are the newly designed precoding matrix set for $K$ users as shown in Eq.~\eqref{Joint_CSI_recon_Pre}, each with $N_t$ antennas and $N_c$ subcarriers.

The entire network is trained in an end-to-end manner. Inputs to the network include the downlink full CSI for the CE phase and downlink data transmission phase, the uplink CSI for the feedback phase, the SNR for each phase and the mask matrix for determining the activated users. The objective of network optimization is to maximize the downlink spectral efficiency, which can be expressed as Eq.~\eqref{Optimization_Objective}. Here, $\Upsilon_k$ denotes the $k$-th element of the mask matrix $\mathbf{\Upsilon} \in \mathbb{R}^{K_{\max}}$. Specifically, $\Upsilon_k=1$ indicates that the corresponding user is activated, while $\Upsilon_k=0$ indicates that the user is inactive. The total number of active users is given by $K=\lVert \mathbf{\Upsilon} \rVert_{1} \leq K_{\max}$. 
% \begin{equation}
% 	\label{R_object}
% 	\mathop {{\rm{maximize}}}\limits_{\theta,\psi, \mathscr{P}} R_{\theta, \psi,\mathscr{P}}.
% \end{equation}

\subsection{Feature Extraction and Dimensional Reconstruction}

By directly extracting the channel CSI for compression at the UE, 
% the source compression rate is indirectly reduced. At the same time, 
both parameter storage and computational overhead are decreased, because the source data for compression is changed from full CSI to a received pilot signal with lower dimensionality. Since the compressed pilot signals have contained the required downlink CSI semantic information of users, a feature extraction network at the BS is used to capture MU CSI semantic features, and a dimensional reconstruction network is employed to reconstruct the feature dimensions. 

\begin{figure}[!t]
	\centering
	\includegraphics[width=3.4in]{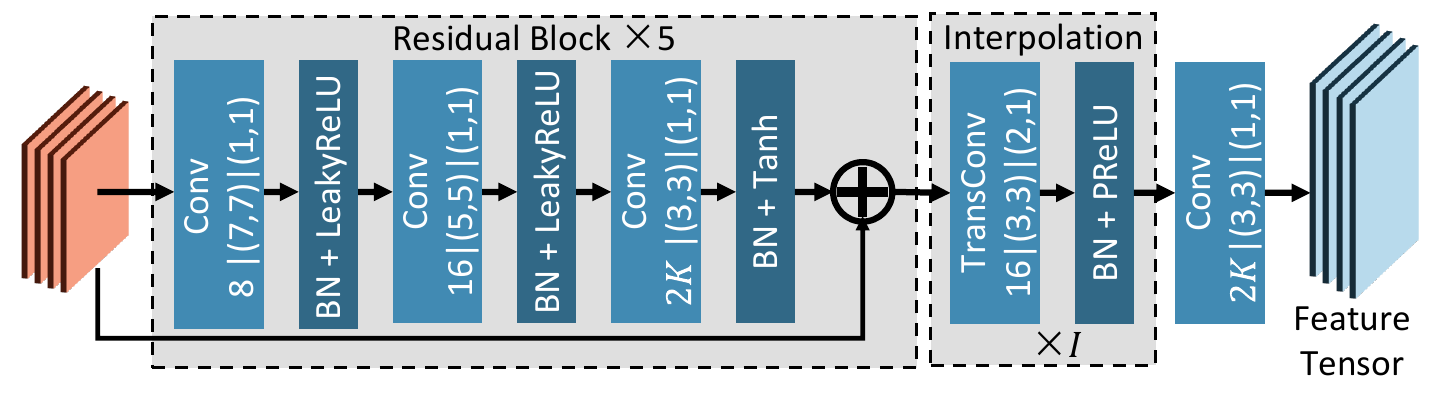}
	\caption{Procession of feature extraction and dimensional reconstruction.}
	\label{FE_UpDim}
\end{figure}

The structure of the feature extraction and dimensional reconstruction networks is shown in Fig.~\ref{FE_UpDim}. First, for the feature tensor of dimension, $M \times N_t \times 2$, features are extracted using multiple residual blocks composed of convolutional layers, BN layers, and activation function layers. The dimensional reconstruction network employs $I$ transposed convolutional layers to reconstruct low-dimensional features to the target dimensionality of the precoding vector by adjusting the stride $S_1$, followed by a convolutional layer to set the number of output feature channels. Overall, the dimensional transformation $M \times N_t \times 2 \rightarrow N_c \times N_t \times 2$ is achieved by the dimensional reconstruction network. The size of $I$ is determined by $M$, $N_c$, and the upsampling step $S_1$, and is typically calculated as $I = N_c / (M S_1)$. If this expression does not yield an integer, $M$ can be adjusted to $M'$ to ensure that the reconstructed feature dimensions remain aligned with $N_c$.

\subsection{Precoding Design}
\begin{figure}[!t]
  \centering
  \subfloat[]{
    \label{PrecodingDesign_Transformer}
    \includegraphics[width=3.3in]{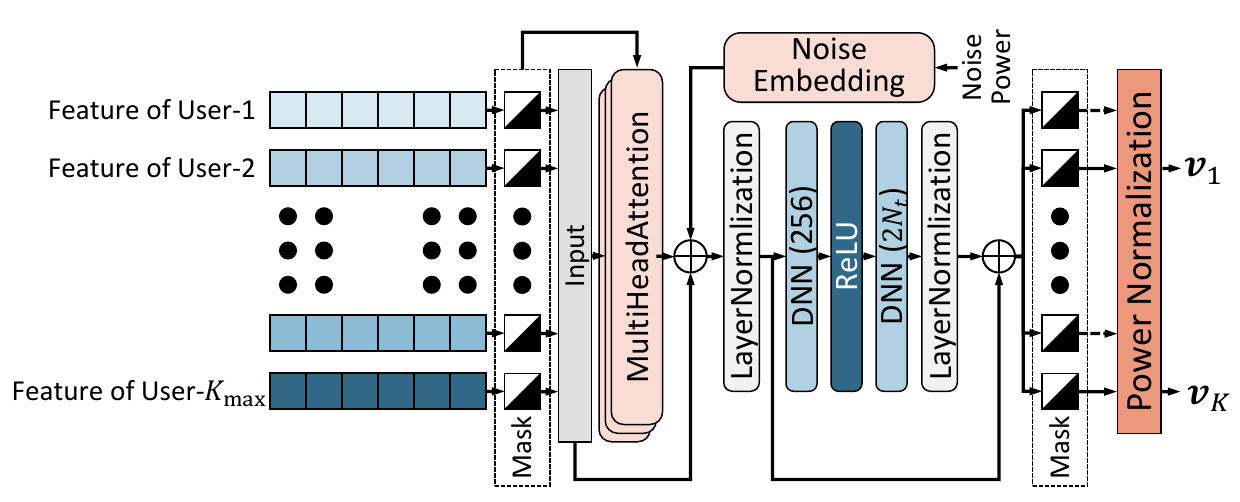}
    
  }\\
  \subfloat[]{
    \label{PrecodingDesign_DNN}
    \includegraphics[width=3.3in]{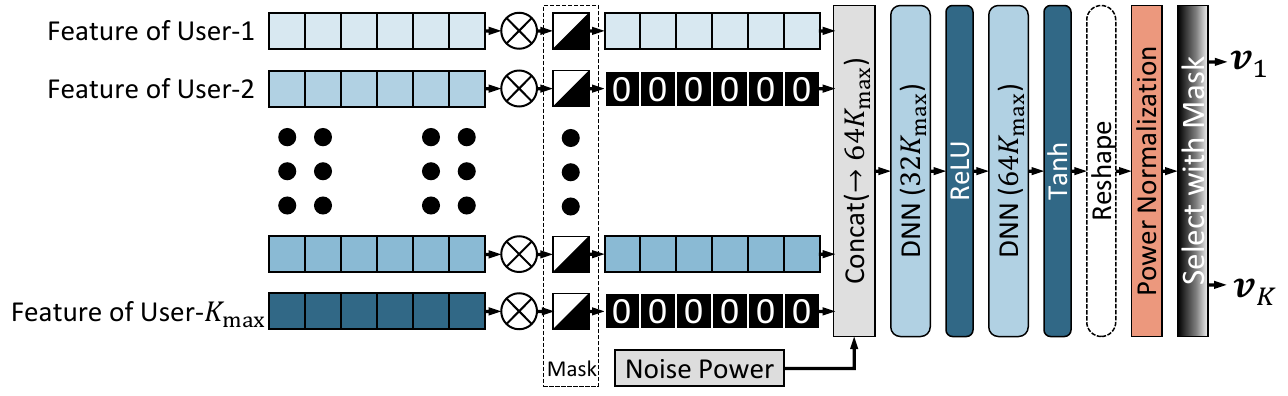}
    
  }\\
  \caption{Architecture of MU precoding design module. \protect\subref{PrecodingDesign_Transformer} Transformer precoding method with proposed user-adaptive strategy. \protect\subref{PrecodingDesign_DNN} DNN precoding method \cite{JFPNet} with zeroing user-adaptive strategy \cite{MU_Adaptive}}.
  \label{Precoder}
\end{figure}

In the precoding design module, we account for the correlation between different subcarriers and employ a consistent network architecture and parameters for the precoding design across subcarriers to reduce both the number of training parameters and storage overhead. 
For the case of a variable number of users, we design a Transformer-based precoding design module utilizing a masked multi-head attention mechanism. The precoding design network for a single subcarrier is also depicted in Fig.~\subref*{PrecodingDesign_Transformer}. Let $\mathbf{X}\in \mathbb{R}^{K_{\max} \times 2N_t}$ denote the features of $K_{\max}$ users on the same subcarrier. The corresponding query, key, and value matrices are expressed as
\begin{equation}
\label{QKV}
\begin{split}
    & \mathbf{Q}^{(h)}=\mathbf{X}\mathbf{W}_{\mathrm{Q}}^{(h)}, 
    \\ &\mathbf{K}^{(h)}=\mathbf{X}\mathbf{W}_{\mathrm{K}}^{(h)}, 
    \\ & \mathbf{V}^{(h)}=\mathbf{X}\mathbf{W}_{\mathrm{V}}^{(h)},
\end{split}
\end{equation}
where $\mathbf{W}_{\mathrm{Q}}^{(h)} \in \mathbb{R}^{2N_t \times E}$, $\mathbf{W}_{\mathrm{K}}^{(h)} \in \mathbb{R}^{2N_t \times E}$, and $\mathbf{W}_{\mathrm{V}}^{(h)} \in \mathbb{R}^{2N_t \times E}$ are trainable parameters associated with the $h$-th attention head. $E$ represent embedding dimension of outputs.

After applying the mask, the output of the $h$-th attention layer can be expressed as
\begin{equation}
    \label{Attention}
    \mathrm{Softmax}\left(\frac{\mathbf{Q}^{(h)}\mathbf{K}^{(h)^T}}{\sqrt{K_{\max}}} \odot \left(\mathbf{\Upsilon}\mathbf{\Upsilon}^T\right)\right)\mathbf{V}^{(h)},
\end{equation}
where $\mathrm{Softmax}(\cdot)$ denotes the softmax function, and $\mathbf{\Upsilon}\mathbf{\Upsilon}^{T}$ represents the mask matrix, ensuring that the attention computation is restricted to the activated users.

In our setup, the embedding dimension $E$ is set to 256. Therefore, the embedding of noise power with shape $\mathbb{R}^{256}$ is incorporated as auxiliary information to enhance precoding design. The output of the multi-head attention layer is added with the input features and the noise power embedding, followed by several FC layers to restore each feature dimension to $2N_t$, corresponding to the real and imaginary parts of the precoding vector. Finally, the mask matrix selects the precoding vectors of the active users, which are normalized to form the final precoding matrix.

Compared with the DNN-based precoding architecture \cite{JFPNet, YuWei_JEFPNet,MU_Adaptive}, which can employs zero-setting operations for user-number adaptation as in~\cite{MU_Adaptive} and is illustrated in Fig.~\subref*{PrecodingDesign_DNN}, our design offers a key advantage: as shown in Eq.~\eqref{QKV}, the number of trainable parameters remains constant regardless of $K_{\max}$. In \cite{MU_Adaptive}, user adaptation is achieved by setting the features of inactive users to zero, which resembles pruning. However, the performance of this approach heavily depends on the depth and width of the following DNN. Specifically, when $K_{\max}$ is large, using a shallow or narrow network results in degraded performance for $K=K_{\max}$, whereas a deep or wide network causes inefficiency and performance loss when $K\ll K_{\max}$. In contrast, the proposed user-adaptive Transformer-based precoding design avoids this trade-off.

When the system is extended to a hybrid beamforming architecture, two precoding design modules should be separately designed for $\mathbf{F}_{\mathrm{RF}}$ and $\mathbf{F}_{\mathrm{BB}}$. It should be noted that the output dimensions of the last layer of the FC for the two networks should be $2N_{\mathrm{RF}}$ and $2N_t$, respectively. Furthermore, for analog precoding, the output must be amplitude-constrained to ensure that only the phase is adjusted.
% The input feature tensor is first concatenated along the user and antenna dimensions, then concatenated with the downlink noise power as known information to assist with power allocation. The concatenated vectors for each subcarrier contain the CSI of MU along with the downlink noise power, so precoding vectors for $K$ users are designed using two FC layers and the tanh activation function, as shown in Fig~\ref{PrecodingDesign}. Note that a power normalization process must follow the last activation function layer to meet the power constraints in Eq.~\eqref{Precoding_constraint}. Finally, the newly designed precoding matrix is reshaped to form a precoding tensor with dimensions: width corresponding to the number of BS antennas $N_t$, length to the number of subcarriers $N_c$, and channels to twice the number of users $K$ (representing real and imaginary components). The set of precoding matrices for $K$ users is then obtained by separating this tensor along the channel dimension, which is expressed as $\mathscr{V}$.

% \begin{figure}[!t]
% 	\centering
% 	\includegraphics[width=3.4in]{PrecodingDesign.pdf}
% 	\caption{Architecture of MU precoding design module. To facilitate understanding, we represent the feature tensor as a set of $N_c \times N_t \times 2K$ squares. Similarly, the precoding tensor is depicted as $N_c \times N_t \times 2K$ squares, with different colors used to distinguish the precoding tensor for different users, each characterized by $N_c \times N_t \times 2$.}
% 	\label{PrecodingDesign}
% \end{figure}

\color{black}
\section{Experimental results}
\label{section3}
%%%%%%%%%%%%%%%%%%%%%%%%%%%%%%%%%
In this paper, we propose an end-to-end JEFPNet with user-number adaptability that integrates three phases of CSI feedback starting at the UE and eventually returning to the UE: downlink CE, uplink CSI feedback, and downlink precoding design. In this section, we demonstrate the effectiveness of the proposed network by comparing our algorithm with traditional algorithms through simulation experiments.
We further validate the effectiveness of the DJSCC architecture for joint CE, CSI feedback, and precoding design, as well as the advantages of this joint design over separated designs. In addition, we demonstrate the effectiveness of the proposed user-number-adaptive mechanism through comparative experiments.
We also verify the performance advantages of the end-to-end multi-module joint designed architecture by comparing the joint network architecture with the corresponding separate network architecture. 
Finally, we evaluate the network’s scalability under varying feedback overheads and the impact of the scenario shift problem on its performance. We also examine the network’s compatibility with existing digital communication systems and provide a statistical characterization of the model storage, distribution, and computational overheads.

This paper considers a MU urban macro-cell (UMa) channel scenario to simulate the proposed network architecture. {In this paper, we adopt a two-sided model similar to \cite{YuWei_JEFPNet, ADJSCC}, where the BS collects channel data during idle periods for offline training and subsequently distributes the trained model. When the UE enters a cell and initiates a communication request, the BS distributes the corresponding encoder model of that cell to the UE.} The users work in an FDD system with an uplink frequency of 2.1 GHz and a downlink frequency of 1.9 GHz. An OFDM system with $N_c=96$ subcarriers and a 10 MHz bandwidth is employed for both uplink and downlink. The BS is equipped with $N_t=32$ logical antenna ports, implemented using $8 \times 8$ dual-polarized physical surface array antennas, with the mapping between logical ports and physical antennas realized through a sparse matrix with power constraints. Each UE is equipped with a single omnidirectional antenna. Other antenna settings, user locations, and scenario configurations follow the 3rd Generation Partnership Project (3GPP) TR 38.901 standard \cite{3GPP38901}. {For example, for the UMa scenario, the cell radius is set to 250~m, the BS height to 25~m, and the minimum UE-to-BS distance to 35~m. The indoor user ratio is set to 80\%.
The channel between the $n_r$-th receiving antenna and the $n_t$-th transmitting antenna can be expressed as}
\begin{equation}
\label{channel}
\begin{split}
    {H_{{n_r},{n_t}}}(\tau ) & = \sqrt {\frac{1}{{K_R + 1}}} H_{{n_r},{n_t}}^{{\rm{NLOS}}}(\tau ) \\ & + \sqrt {\frac{K_R}{{K_R + 1}}} H_{{n_r},{n_t}}^{{\rm{LOS}}}(\tau )\delta (\tau  - {\tau _1})
\end{split}
\end{equation}
{where $\delta(\cdot)$ denotes the Dirac delta function, $K_R$ is the Rician $K$-factor determined by the propagation scenario, $T$ represents the cluster delays, $H_{{n_r},{n_t}}^{{\rm{LOS}}}$ is the line-of-sight (LOS) channel component, and $H_{{n_r},{n_t}}^{{\rm{NLOS}}}$ is the sum of the non-line-of-sight (NLOS) multipath components.}

In the downlink CE phase, sparse pilots with equal spacing are utilized, with a pilot interval of $g=4$, corresponding to $M=24$. User locations are randomly distributed following the standard, and QuaDriGa \cite{QuaDriGa,QuaDriGa_3D} is used to generate the corresponding uplink and downlink CSI for each user. 
% \textcolor{blue}{For the UMa scenario, the QuaDriGa configuration is set to ``\texttt{3GPP\_38.901\_UMa}'', with the default parameters applied, including 17 channel clusters.}
For the UMa scenario, the QuaDriGa configuration is set to ``\texttt{3GPP\_38.901\_UMa}'', with the default parameters applied, including 17 channel clusters.
The uplink and downlink CSI of $K$ users are aggregated to form one sample. The dataset comprises 100,000 training samples, 30,000 validation samples, and 10,000 test samples. {In practical deployments, datasets can be acquired through idle-time measurements, and the network can be trained offline.} Training is conducted over 500 epochs, utilizing the Adam optimizer with a batch size of 128. The initial learning rate is set to 0.001 and is reduced by half if the validation loss does not improve for 20 consecutive epochs. The downlink spectral efficiency is calculated 
 with Eq.~\eqref{Sum_rate} under a downlink power constraint of $P = 1$.

\subsection{Separate Architecture}
\begin{figure*}[!t]
	\centering
	\includegraphics[width=6.0in]{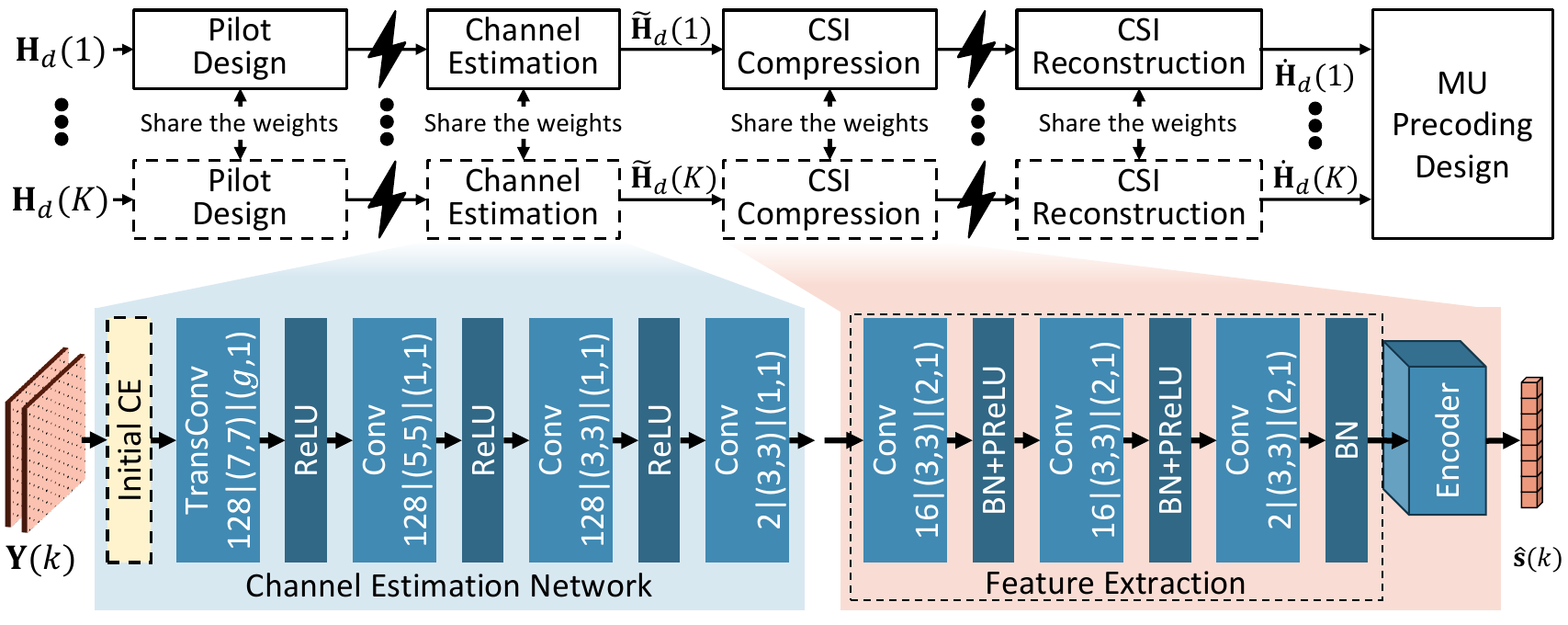}
	\caption{The system model includes pilot design, CE, CSI feedback, and precoding design within a separate architecture framework. Additionally, network architecture diagrams for some modules not featured in the joint architecture are provided. For different users, the parameters are shared across modules because they operate in the same scatterer environment.}
	\label{SEFPNet_UE}
\end{figure*}
To evaluate the performance differences between the joint and separate architectures, we designed a separate architecture comprising downlink CE, CSI compression, CSI reconstruction, and precoding design tasks, as shown in Fig.~\ref{SEFPNet_UE}. At the UE, instead of the joint CE and CSI compression used in JEFPNet, we first employ a CE network to estimate the downlink CSI. Next, a feature extraction network is applied to extract the downlink CSI features in the frequency domain, followed by CSI compression using the same architecture depicted in Fig.~\ref{Encoder}. The compressed CSI is then transmitted. The details of the CE network and the processing of downlink CSI feature extraction are both illustrated in Fig.~\ref{SEFPNet_UE}. {The parameter settings of the CE module follow those reported in existing CE studies \cite{CE_Methods}, while the feature extraction network is adapted from the analysis transformation network used in CSI feedback research \cite{ADJSCC}.} For the CE network, we begin by using the learnable parameter matrix $\mathbf{O}_{\mathcal{M}_m} \in \mathbb{C}^{L \times N_t}$ with the bias vector $\mathbf{b}_{\mathcal{M}_m} \in \mathbb{C}^{N_t}$ for each subcarrier posted pilots, i.e., $\mathcal{M}_m \in \mathcal{M}$ to simulate the linear minimum mean square error (LMMSE) CE, providing the initial estimation result \cite{CE_Methods}. Similar to the operation of the learnable pilot matrix, the real and imaginary parts of the received pilot signal $\mathbf{Y}(k)$ of the $K$-th user are processed separately, i.e., the real part of the initial estimated downlink CSI of the $\mathcal{M}_m$-th subcarrier can be expressed as $\mathcal{R}\big({\mathbf{y}_{\mathcal{M}_m}(k)}\big)\mathcal{R}\big({\mathbf{O}_{\mathcal{M}_m}}\big)- \mathcal{I}\big({\mathbf{y}_{\mathcal{M}_m}(k)}\big)\mathcal{I}\big({\mathbf{O}_{\mathcal{M}_m}}\big)+\mathcal{R}\big({\mathbf{b}_{\mathcal{M}_m}}\big)$, and the imaginary part can be expressed as $\mathcal{R}\big({\mathbf{y}_{\mathcal{M}_m}(k)}\big)\mathcal{I}\big({\mathbf{O}_{\mathcal{M}_m}}\big)+\mathcal{I}\big({\mathbf{y}_{\mathcal{M}_m}(k)}\big)\mathcal{R}\big({\mathbf{O}_{\mathcal{M}_m}}\big)+\mathcal{I}\big({\mathbf{b}_{\mathcal{M}_m}}\big)$. A transpose convolution layer is then employed to perform upsampling in the frequency domain, followed by several convolution layers and ReLU activation layers for feature extraction and channel reconstruction. As illustrated in Fig.~\ref{SEFPNet_UE}, the feature extraction process consists of convolution and transpose convolution layers, along with batch normalization layers and activation functions.

At the BS, the downlink CSI is first reconstructed using a CSI reconstruction network, which employs an architecture similar to the decoder in \cite{ADJSCC}. The CSI of different users is processed using CSI reconstruction networks with the same parameters. Following CSI reconstruction, precoding vectors are designed based on the reconstructed CSI of the MU.

It is important to note that in the separate architecture, the modules performing distinct tasks are individually trained with optimization objectives tailored to their respective tasks and then integrated for end-to-end performance testing. For the downlink CE task, a pilot design and CE network is utilized, and its loss function aims to maximize estimation accuracy, represented by
\begin{equation}
	\label{LOSS_CE}
	\mathcal{L}_{CE} = \mathrm{minimize}\left\{{\mathrm{MSE}\left({\mathbf{H}_{d}(k), \tilde{\mathbf{H}}_{d}(k)}\right)}\right\},
\end{equation}
where $\mathrm{MSE}\left(\cdot\right)$ denotes the mean square error (MSE) between the two inputs. $\tilde{\mathbf{H}}_{d}(k)$ is the estimated downlink CSI of the $k$-th user as shown in Fig.~\ref{SEFPNet_UE}

Similarly, the CSI compression and reconstruction network is employed for the CSI feedback task, and its training loss function maximizes reconstruction accuracy, which also can be expressed as
\begin{equation}
	\label{LOSS_FB}
	\mathcal{L}_{FB} = \mathrm{minimize}\left\{{\mathrm{MSE}\left({\mathbf{H}_{d}(k), \dot{\mathbf{H}}_{d}(k)}\right)}\right\},
\end{equation}
where $\dot{\mathbf{H}}_{d}(k)$ is the reconstructed downlink full CSI of the $k$-th user as shown in Fig.~\ref{SEFPNet_UE}.

For the downlink precoding design processing, we consider two approaches: traditional precoding design and AI-based precoding design. {In the traditional scheme, we employ the singular value decomposition (SVD) and minimum mean squared error (MMSE) precoding methods, both of which utilize the outputs $\dot{\mathbf{H}}_{d}(1)$ to $\dot{\mathbf{H}}_{d}(K)$ from the CSI reconstruction module to design the precoding vectors for $K$ users.} The SVD algorithm is applied to different subcarriers to obtain the right singular vectors and eigenvalues for the $K$ users. These right singular vectors are then used as the precoding vectors for transmitting the downlink data. {Ideally, the precoding matrix obtained by the MMSE algorithm for the $n_c$-th subcarrier can be expressed as \cite{MMSE_1, MMSE_2}}
{
\begin{equation}
	\label{MMSE}
	\mathbf{V}_{n_c}=\mathbf{H}_{n_c}\left({\mathbf{H}_{n_c}^H\mathbf{H}_{n_c}+\frac{K\sigma_{d}^2}{P}\mathbf{I}}\right)^{-1},
\end{equation}
where $\mathbf{H}_{n_c}=\left[{\mathbf{h}_{d, n_c}^{T}(1),\cdots, \mathbf{h}_{d, n_c}^{T}(K)}\right]^T\in \mathbb{C}^{K \times N_t}$ is MU downlink channel matrix. Under non-ideal case, the design of $\mathbf{V}_{n_c}$ must rely on the reconstructed CSI, i.e., from $\dot{\mathbf{H}}_{d}(1)$ to $\dot{\mathbf{H}}_{d}(K)$.}
Downlink power allocation is performed using the water-filling (WF) algorithm, which utilizes the eigenvalues and the downlink noise power $\sigma^2_{d}$ to determine the downlink transmission power for the $K$ users. For the AI-based precoding design, we use the network architecture ideas as \cite{JFPNet} to jointly design the CSI feedback and precoding design networks. Both the CSI feedback module and the precoding design module share the same network architecture as JEFPNet to ensure a fair comparison. This approach is referred to as ``JFPNet,'' and its loss function is defined as maximizing the downlink spectral efficiency, using the ideally estimated CSI as the source for joint CSI feedback and precoding design.

\subsection{Performance Comparison Among Different Architectures}
\begin{table}[!t]
\renewcommand{\arraystretch}{1.15}
\centering
\caption{Impact of CE and CSI feedback errors on Traditional precoding performance}
\begin{tabular}{c|c|ccccc}
\toprule
\textbf{Algorithm} & $\bm{\mathrm{SNR_{u}}}$ & \textbf{Ideal} & \textbf{Ideal FB} & \textbf{Ideal CE} & \textbf{Non-Ideal} \\ 
\midrule
\multirow{3}{*}{\textbf{SVD}} 
& -10 dB & 14.405 & 10.803 & 8.282 & 7.931 \\ 
& 0 dB   & 14.405 & 10.803 & 9.581 &  9.119 \\ 
& 10 dB  & 14.405 & 10.803 & 9.747 & 9.265 \\ 
 \cmidrule{1-6}
\multirow{3}{*}{\textbf{MMSE}} 
& -10 dB & 14.32 & 11.162 & 8.717 & 8.396 \\ 
& 0 dB   & 14.405 & 11.162 & 9.925 & 9.514 \\ 
& 10 dB  & 14.405 & 11.162 & 10.077 & 9.653 \\ 
\toprule
\end{tabular}
\label{Traditional_Precoder}
\end{table}

\begin{figure}[!t]
	\centering
	\includegraphics[width=3.4in]{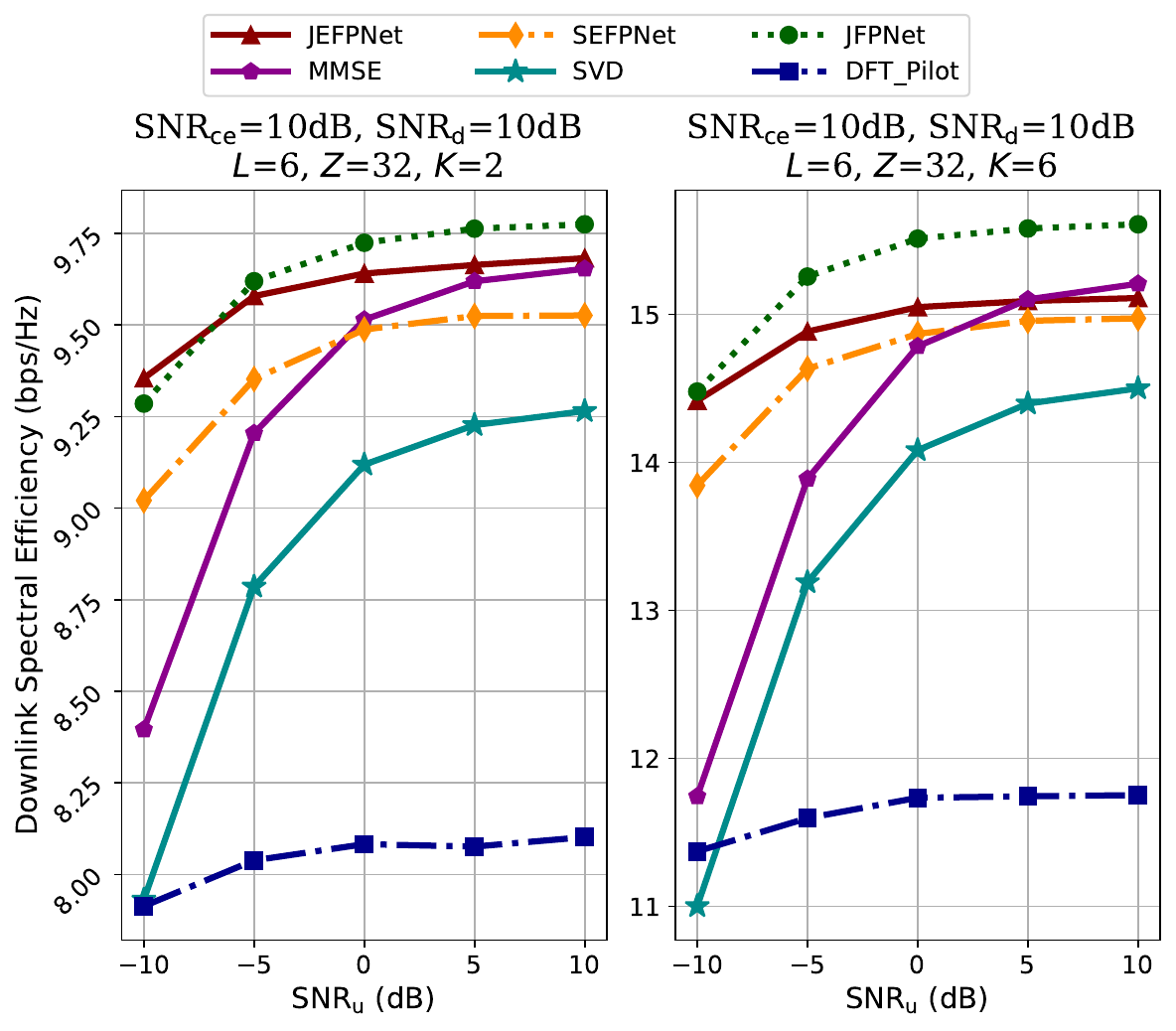}
	\caption{Performance comparison of JEFPNet with traditional separate architecture and partial separate architecture for CE, feedback, and precoding design.}
	\label{2_Overall_gain}
\end{figure}

% \begin{table}[!ht]
% \centering
%     \caption{Impact of CE and CSI FB accuracy on the performance of conventional SVD precoding algorithms}
% \begin{tabular}{|c|c|c|c|c|}
% \hline
% $Z$                 & $\mathrm{SNR_{u}}$ & Ideal CE \& FB & Ideal CE & SVD \& WF \\ \hline
% \multirow{5}{*}{32} & -10 dB             & 14.405         & 3.027    & 3.006     \\ \cline{2-5} 
%                     & -5 dB              & 14.405         & 3.184    & 3.140     \\ \cline{2-5} 
%                     & 0 dB               & 14.405         & 3.244    & 3.200     \\ \cline{2-5} 
%                     & 5 dB               & 14.405         & 3.267    & 3.217     \\ \cline{2-5} 
%                     & 10 dB              & 14.405         & 3.274    & 3.244     \\ \hline
% \multirow{5}{*}{96} & -10 dB             & 14.405         & 3.285    & 3.232     \\ \cline{2-5} 
%                     & -5 dB              & 14.405         & 3.424    & 3.363     \\ \cline{2-5} 
%                     & 0 dB               & 14.405         & 3.470    & 3.410     \\ \cline{2-5} 
%                     & 5 dB               & 14.405         & 3.487    & 3.425     \\ \cline{2-5} 
%                     & 10 dB              & 14.405         & 3.493    & 3.431     \\ \hline
% \end{tabular}
% \end{table}
In this section, we compare the downlink spectral efficiencies of three different architectures: JEFPNet in a joint architecture, SVD precoding and MMSE precoding with WF power allocation in a separate architecture, and JFPNet in a partially separate architecture. The experiments are carried out with the $\mathrm{SNR_{ce}}$ in the CE phase and the $\mathrm{SNR_{d}}$ in the downlink data transmission phase fixed. The number of CE symbols ($L$) is set to 6, the feedback overhead ($Z$) is 32, and the maximum number of users ($K_{\mathrm{max}}$) is 6. By varying the SNR of the uplink feedback channel, i.e. $\mathrm{SNR_{u}}$, we evaluate the downlink spectral efficiency for each architecture. We evaluate the network performance in Fig.~\ref{2_Overall_gain} for the cases $K=2$ and $K=6$. The curve labeled ``JFPNet'' depicts the JFPNet network architecture under ideal CE. The curve ``SEFPNet'' indicates the performance when the CE module is separately trained with JFPNet and tested together after concatenation. The ``SVD'' and ``MMSE'' curves represent the performance of the separate architecture, where the CE and CSI feedback modules are separately trained, and then concatenated for testing to generate reconstructed downlink CSI for precoding design with SVD and MMSE algorithms.

The experimental results show that, {for traditional algorithms, the errors introduced by CE and CSI feedback have a smaller impact on MMSE than on SVD. Compared with these traditional approaches, the proposed JEFPNet consistently achieves near-optimal performance, particularly at low $\mathrm{SNR_u}$. For example, with $K=2$ and $\mathrm{SNR_u}=-10$~dB, JEFPNet delivers performance gains of 1.42~bps/Hz over SVD and 0.96~bps/Hz over MMSE. Although MMSE slightly outperforms JEFPNet by 0.1~bps/Hz when $K=6$ and $\mathrm{SNR_u}=10$~dB, this does not undermine the performance advantage and effectiveness of the proposed network. This result can be attributed to the fact that JEFPNet must balance the overall average performance across different $\mathrm{SNR_u}$ and $K$ values simultaneously. Consequently, it prioritizes improving performance at low $\mathrm{SNR_u}$ and large $K$, as these scenarios exert a greater impact on the average performance. For instance, at $K=2$, JEFPNet achieves a gain of 0.956~bps/Hz over MMSE at $\mathrm{SNR_u}=-10$~dB, while the gain is only 0.029~bps/Hz at $\mathrm{SNR_u}=10$~dB. Furthermore, at $\mathrm{SNR_u}=-10$~dB and $K=6$, JEFPNet provides a gain of 2.67~bps/Hz compared to MMSE, and 0.959~bps/Hz at $K=2$.} In the case of $K=2$, compared to the ``SEFPNet'', which also accounts for CE and CSI feedback errors, JEFPNet achieves a performance gain of approximately 0.33 bps / Hz at $\mathrm{SNR_{u}}= -10$ dB and 0.17 bps/Hz at $\mathrm{SNR_{u}}= 10$ dB. 

{To evaluate the significance of the pilot design module, we replace the AI-based pilot design module in JEFPNet with a traditional orthogonal discrete cosine transform (DFT)-based pilot design \cite{CE_Methods}, while keeping the rest of the architecture unchanged and continuing to train the network in an end-to-end manner. The results, shown by the ``DFT\_Pilot'' curves in Fig.~\ref{2_Overall_gain}, reveal a substantial performance degradation compared with JEFPNet. Specifically, the performance decreases by 1.44\~1.58~bps/Hz for $K=2$ and by approximately 3.05\~3.39~bps/Hz for $K=6$. This degradation occurs because, under limited pilot resources, joint optimization of the pilot design module with other modules allows for the retention of more feature information relevant to the precoding task. However, the fixed pilot can't effectively save feature information.}

% Additionally, even when compared to ``JFPNet,'' which utilizes the ideal downlink full CSI as its source, the JEFPNet architecture still demonstrates a performance gain of 0.07 bps/Hz at $\mathrm{SNR_{u}}=-10$ dB. This improvement can be attributed to the joint estimation and feedback architecture employed by JEFPNet. Although both ``SEFPNet'' and ``JFPNet'' utilize CSI compression networks similar to JEFPNet, they compress the source using full CSI, while JEFPNet compresses the source using a lower-dimensional pilot signal instead. As a result, JEFPNet equivalently reduces the compression rate for the same output dimension. Meanwhile, with the assumption of limited transmission resources, directly extracting effective features from the received pilot signals with an end-to-end training manner with specific task objectives can enhance the efficiency of transmitting relevant semantic information, which in turn improves the downlink spectral efficiency. 

To investigate the performance loss of traditional algorithms, we compare the effects of CE errors and CSI feedback errors on traditional precoding design, with the results summarized in Table~\ref{Traditional_Precoder}. The experimental configurations in Table~\ref {Traditional_Precoder} are identical to those in Fig.~\ref {2_Overall_gain}, except that we have set the evaluated $K$ to be 2. In Table~\ref{Traditional_Precoder}, ``Ideal'' refers to the case that traditional precoding and the WF algorithm are used with ideal CE and CSI feedback. ``Ideal FB'' represents the case where erroneous CE is used with ideal CSI feedback, alongside the traditional precoding method. ``Ideal CE'' indicates the case where only the effect of CSI feedback on the traditional precoding algorithm is considered. ``Non-Ideal'' denotes that errors from CE and CSI feedback are taken into account when employing traditional precoding algorithms. The results in Table~\ref{Traditional_Precoder} show that the CSI feedback errors have a significant impact on the performance of the traditional algorithm. Specifically, when the feedback overhead $Z$ is 32 and the feedback channel quality $\mathrm{SNR_{u}}$ is -10 dB, for the SVD precoding algorithm, CE errors lead to a performance degradation of 3.602~bps/Hz, while CSI feedback errors lead to a performance degradation of approximately 6.123~bps/Hz. Moreover, with the existence of CSI feedback errors, the addition of CE errors results in a performance loss of 0.351~bps/Hz. The smaller performance loss caused by CE error can be attributed to the large CE overhead ($L$) and the high $\mathrm{SNR_{ce}}$ of the downlink CE, both of which mitigate its impact, while the CSI feedback overhead remains relatively small. Nevertheless, the joint design of CSI feedback and precoding significantly reduces this performance loss, as demonstrated by the experimental results in Fig.~\ref{2_Overall_gain} with curves ``SEFPNet'' and ``JEFPNet''. {For the MMSE precoding algorithm, under the same conditions, although the performance is 0.085~bps/Hz lower than that of the SVD algorithm in the ``Ideal'' case, the performance degradation is limited to 5.924~bps/Hz with the errors, which remains 0.46~bps/Hz higher than that of the SVD algorithm.}

% By comparing ``SEFPNet'' with ``SVD \& WF'' in Fig.~\ref{2_Overall_gain}, we observe that the performance impact is not significant due to the low CE error, which is attributed to the large CE overhead ($L$) and the high $\mathrm{SNR_{ce}}$ of the downlink channel. The performance loss for traditional precoding at $\mathrm{SNR_{u}}=-10 \mathrm{ dB}$ is only 0.273 bps/Hz, while the AI-based precoding scheme in the partially separated architecture results in a loss of approximately 0.273 bps/Hz. 

\begin{figure}[!t]
	\centering
	\includegraphics[width=3.4in]{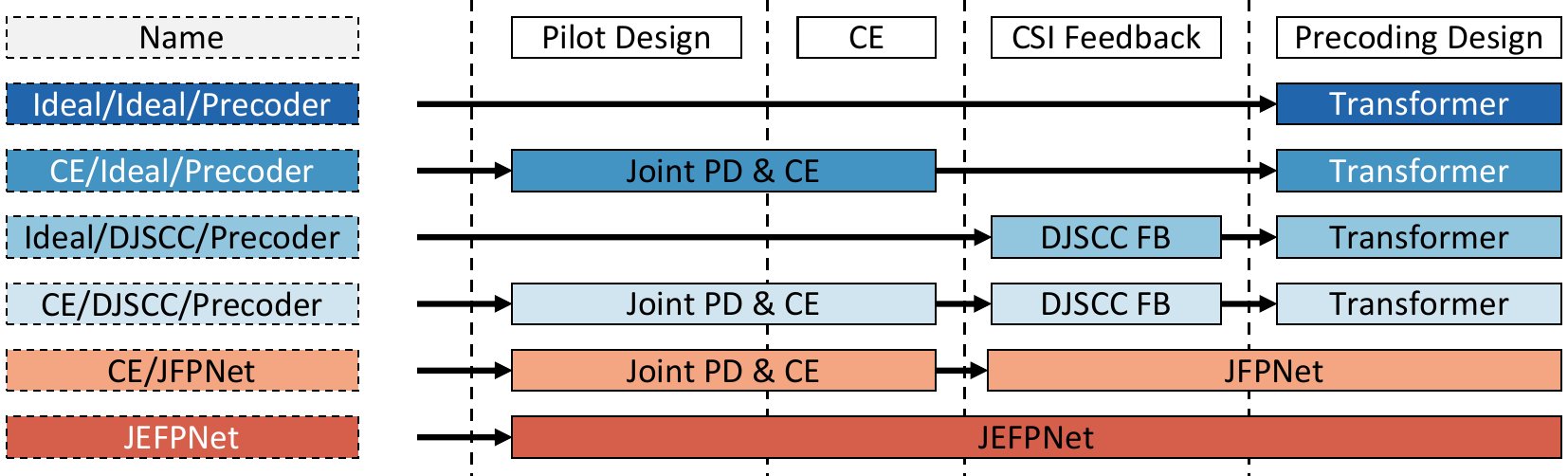}
	\caption{Network components. A direct connection between the regions corresponding to a specific module indicates that this module is assumed to be ideal. ``Transformer'' denotes the precoding design module in JEFPNet, which is trained separately; ``Joint PD \& CE'' denotes the joint pilot design and CE network in SEFPNet; and ``DJSCC FB'' denotes the CSI feedback network based on the DJSCC architecture.}
	\label{1_ablation}
\end{figure}

\begin{figure}[!t]
	\centering
	\includegraphics[width=3.4in]{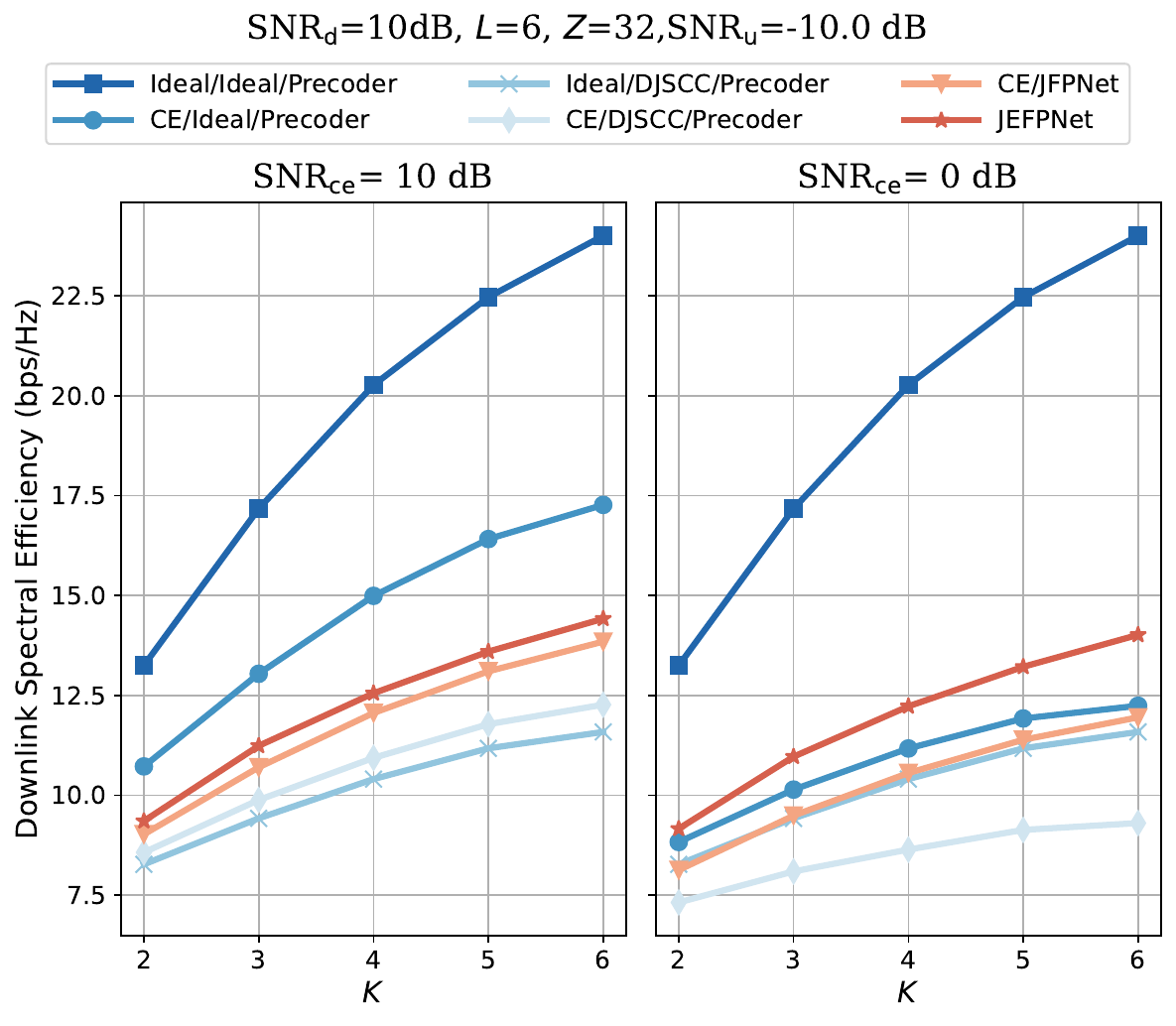}
	\caption{Ablation experiment.}
	\label{1_ablation_EX}
\end{figure}

{To assess the impact of each module on system performance, we examine the effect of introducing different AI modules relative to the ideal case under the conditions $L = 6$, $Z = 32$, $\mathrm{SNR}_u = -10$\,dB, $\mathrm{SNR}_{\mathrm{ce}} \in \{10~\text{dB}, 0~\text{dB}\}$. Six scenarios are considered, as illustrated in Fig.~\ref{1_ablation}. The experimental results are shown in the Fig.~\ref{1_ablation_EX}}

{By comparing the performance of ``CE/Ideal/Precoder'' and ``JEFPNet'' at $\mathrm{SNR}_{\mathrm{CE}} = 10$~dB and $0$~dB, we observe that CE errors degrade the performance of the separated architecture more severely than that of the joint architecture. Specifically, at $K = 6$ and $\mathrm{SNR}_{\mathrm{ce}} = 0$ dB, CE error causes a performance loss of 11.76~bps/Hz, feedback error causes a loss of 12.42~bps/Hz, and their combined effect results in a loss of 14.70~bps/Hz. Compared with the separated architecture ``CE/DJSCC/Precoder,'' joint feedback and precoding design improves performance by 2.65~bps/Hz, whereas joint CE-feedback-precoding design achieves an improvement of 4.71~bps/Hz.}

\subsection{Adaptive Testing of the Number of Users}
\begin{figure}[!t]
	\centering
	\includegraphics[width=3.4in]{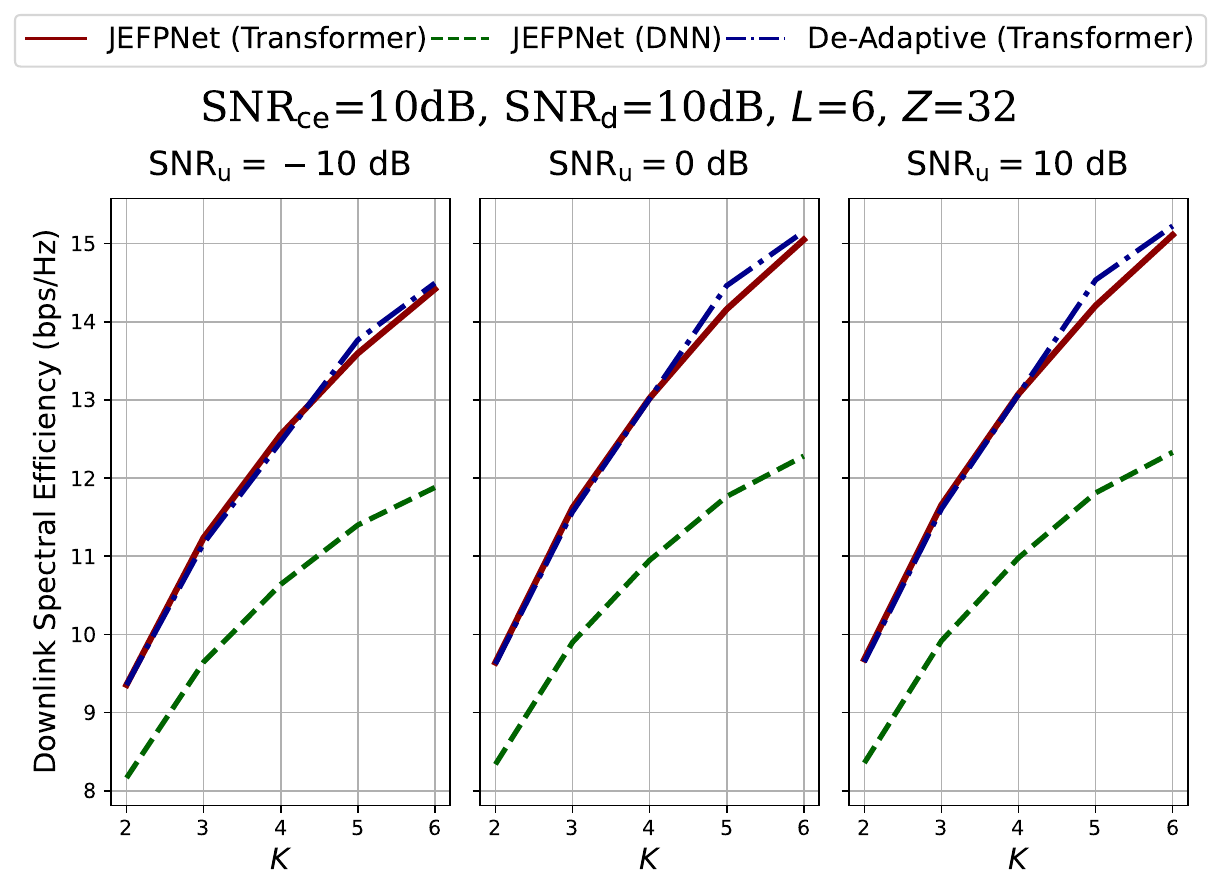}
	\caption{Adaptive Testing of the Number of Users. $K_max$ is setting 6 for architectures with user-number adaptability.}
	\label{3_User_Adaptive}
\end{figure}
To demonstrate the effectiveness of the proposed architecture, we evaluate the performance of JEFPNet under both a fixed and a variable number of users, referred to as ``JEFPNet (Transformer)'' and ``De-Adaptive (Transformer)'', respectively. At the same time, we replace the proposed precoding network in JEFPNet (Fig.~\subref*{PrecodingDesign_Transformer}) with a user-adaptive precoding network with user-adaptive machine in \cite{MU_Adaptive} (Fig.~\subref*{PrecodingDesign_DNN}), referred to as ``JEFPNet (DNN)''. Notably, ``JEFPNet (Transformer)'' requires 258,300 learnable parameters on the BS side, whereas ``De-Adaptive (Transformer)'' must store separate models for each user configuration, each comprising 258,300 parameters. The ``JEFPNet (DNN)'' architecture requires 273,148 learnable parameters. The remaining parameter settings are provided in the caption of Fig.~\ref{3_User_Adaptive}. Experimental results demonstrate that ``JEFPNet (Transformer)'' outperforms ``JEFPNet (DNN)'', particularly as the number of users increases. Specifically, at $\mathrm{SNR}_{u}=10$~dB and $K=6$, ``JEFPNet (Transformer) achieves a performance gain of 2.78~bps/Hz. Furthermore, the performance of ``JEFPNet (Transformer)'' with user-number adaptability is comparable to that of ``De-Adaptive (Transformer)'', with only a 0.11~bps/Hz performance gap under the same conditions. 

By comparing ``JEFPNet (Transformer)’’ and ``De-Adaptive (Transformer),’’ we demonstrate that the proposed network achieves performance comparable to that of a network trained with a fixed number of users, while reducing network storage overhead by a factor of $1/K_{\max}$. Furthermore, the user-number adaptability design of JEFPNet eliminates the need for model switching when the number of users changes, making it particularly suitable for two-sided architectures where encoder models must be distributed to UEs from BS.

\color{black}
\subsection{Evaluating the Validity of Joint CE within the DJSCC Framework}
\begin{figure}[!t]
	\centering
	\includegraphics[width=3.4in]{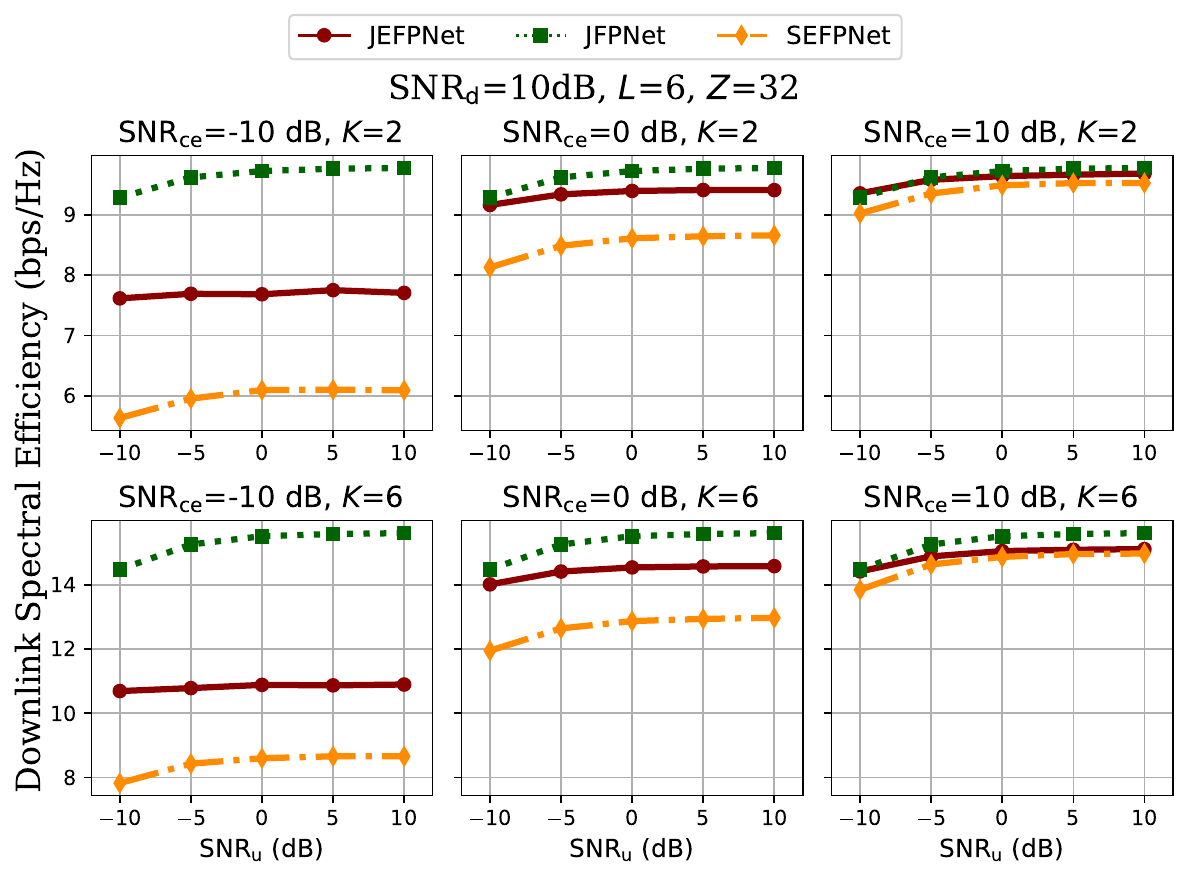}
	\caption{Impact of different CE errors on the performance of separate and joint architectures.}
	\label{4_Diff_SNRce}
\end{figure}

\begin{figure}[!t]
	\centering
	\includegraphics[width=3.4in]{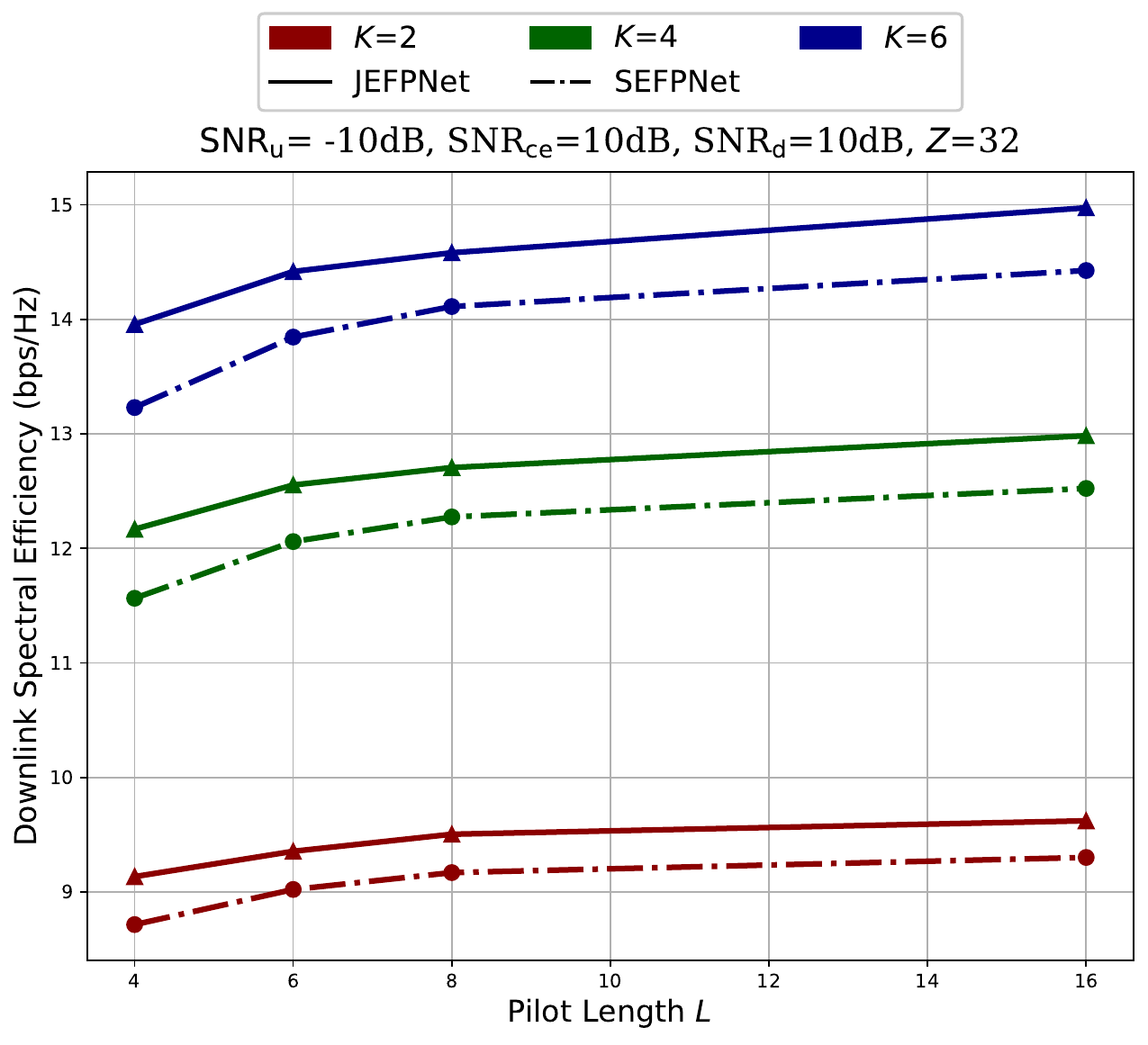}
	\caption{Impact of different CE overheads on the performance of separate and joint architectures.}
	\label{5_Diff_L}
\end{figure}

JEFPNet is trained in an end-to-end manner that incorporates CE errors into the optimization of network parameters. In this subsection, we investigate the impact of estimation overhead and the channel quality during the CE phase on the final performance metrics (downlink spectral efficiency) of the network. In the experiment corresponding to Fig.~\ref{4_Diff_SNRce}, we fix $L$, set $K_{\max}=6$, $Z=32$, and $L=6$, and evaluate the impact of CE error rate on performance. Specifically, we compare JEFPNet with a joint architecture and SEFPNet with a separate architecture under $K\in\{2,6\}$ by varying $\mathrm{SNR_{ce}}$ at $-10$, $0$, and $10$~dB. {In the experiment corresponding to Fig.~\ref{5_Diff_L}, we fix $\mathrm{SNR_{ce}}=10$~dB and $\mathrm{SNR_{u}}=-10$~dB, set $L\in\{2,4,6,16\}$, and keep other parameters consistent with those in Fig.~\ref{4_Diff_SNRce}. We then assess the impact of pilot length, which is also called CE overhead in this paper, on the performance of JEFPNet and SEFPNet under $K\in\{2,4,6\}$.}

From the experimental results in Fig.~\ref{4_Diff_SNRce}, we observe that as the channel SNR decreases, the performance of both the separate and joint architectures deteriorates. However, the performance of JEFPNet under the joint architecture consistently outperforms SEFPNet under the separate architecture. When $\mathrm{SNR_{ce}}$ is reduced from 10 dB to 0 dB, the performance of JEFPNet decreases by only 0.193 bps/Hz at $\mathrm{SNR_{u}} = -10$ dB for the $L=6$ case, while SEFPNet experiences a greater drop of 0.894 bps/Hz under the same conditions. As $\mathrm{SNR_{ce}}$ further decreases from 10 dB to -10 dB, the difference between the two architectures becomes even more pronounced. The performance of JEFPNet decreases by 1.74 bps/Hz, while SEFPNet’s performance significantly decreases by 3.39 bps/Hz. The above phenomena can be attributed to the fact that SEFPNet uses ground-truth CSI for training. When the channel quality during estimation is poor, the CSI fed back during testing deviates more from the ground-truth CSI distribution, leading to degraded performance in the subsequent joint feedback and precoding design module.

{We validate the effectiveness of CE within the DJSCC framework by evaluating SEFPNet with a separate CE design and JEFPNet with a joint CE and CSI feedback design.
% Since JFPNet is independent of CE, its performance remains unchanged as $L$ varies. Meanwhile, i
Increasing the CE overhead $L$ improves downlink spectral efficiency of JEFPNet and SEFPNet, as shown in Fig.~\ref{5_Diff_L}. This trend arises because a larger $L$ enables JEFPNet to extract richer semantic information from the pilot signals received at the UE, while allowing SEFPNet to achieve higher CE accuracy and thereby decrease the impact of estimation errors.
% Moreover, as $L$ increases, the performance of SEFPNet gradually approaches that of JFPNet under ideal CE, while JEFPNet even surpasses JFPNet. 
Specifically, JEFPNet outperforms JFPNet by 0.495~bps/Hz for $L=16$, $K=6$, and by 0.335~bps/Hz for $K=2$. Furthermore, as $L$ increases, the performance gap between JEFPNet and SEFPNet gradually narrows. For example, when $K = 6$ under the same conditions, the gap is $0.419$ bps/Hz at $L = 4$ and decreases to $0.320$ bps/Hz at $L = 16$. }

\subsection{Analysis of Storage and Computational Complexity Overhead}
In this subsection, we analyze the training parameters and floating point operations (FLOPs) of different architectures under specific configurations: the number of feedbacks $Z=32$, the number of subcarriers $N_c=96$, the CE overhead $L=6$, the pilot interval $g=4$, and the number of users $K_{\max}=6$.

\begin{figure}[!t]
  \centering
  \label{FLOPs_Parameters}
  \subfloat[]{
    \label{UE_FLOPs}
    \includegraphics[width=3.3in]{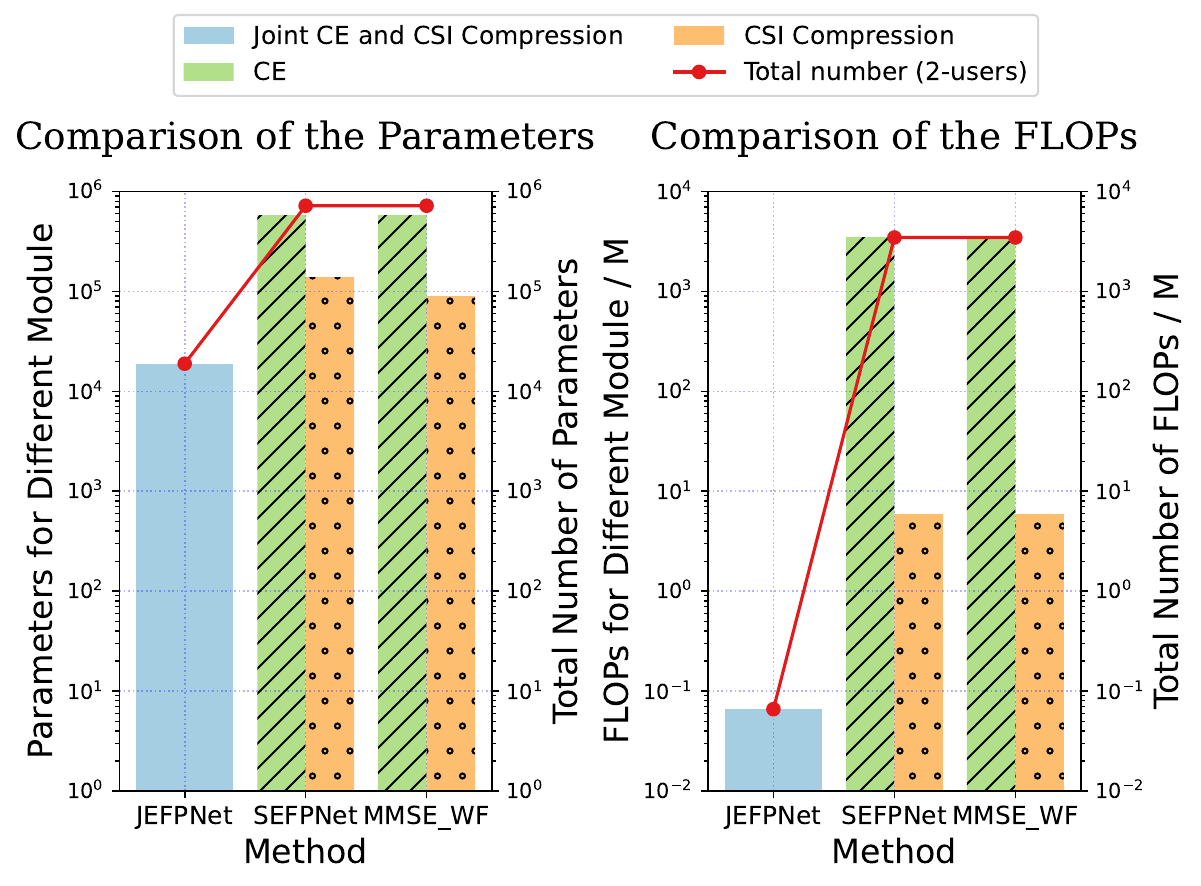}
    
  }\\
  \subfloat[]{
    \label{BS_FLOPs}
    \includegraphics[width=3.3in]{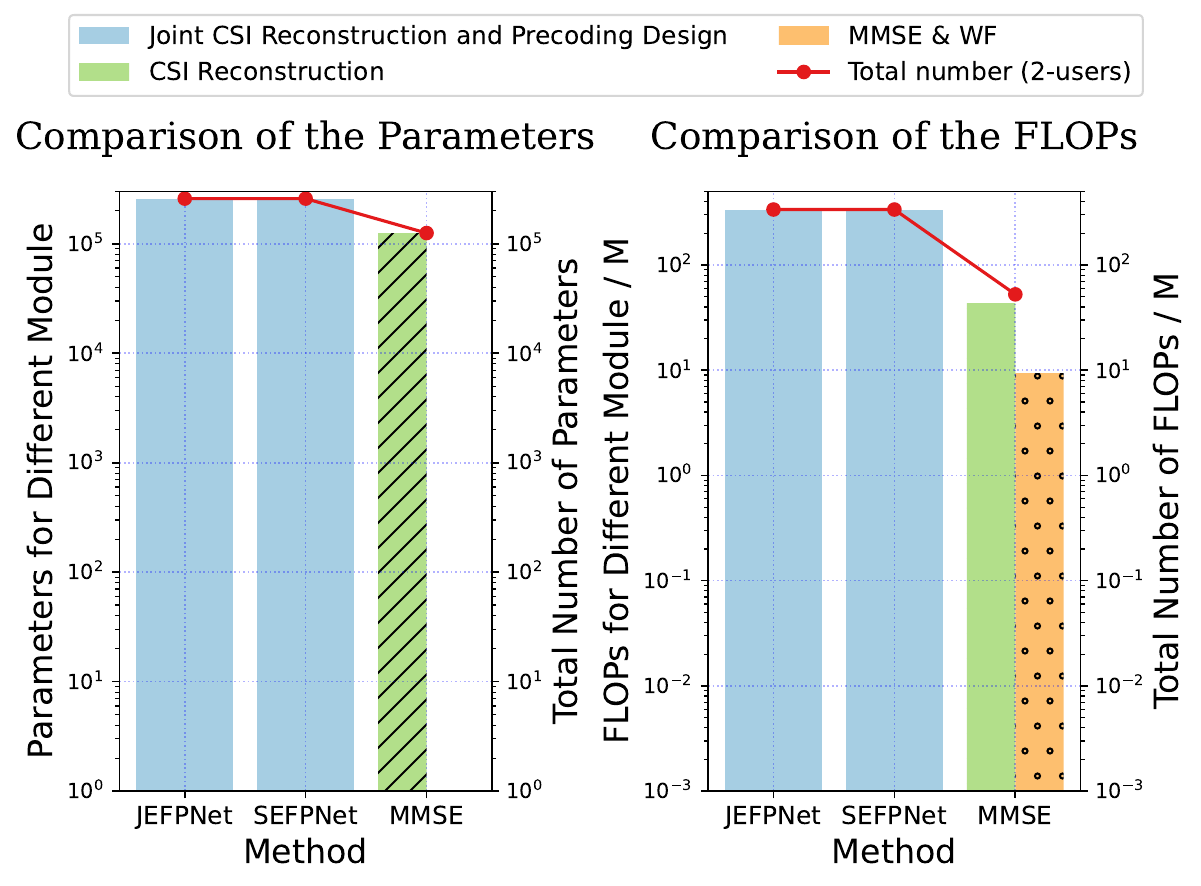}
    
  }\\
  \caption{Statistics on storage overhead and computational complexity for both sides of the communication system with $K=2$, considering different network architectures: \protect\subref{UE_FLOPs} the UE side; \protect\subref{BS_FLOPs} the BS side.}
\end{figure}

Fig.~\subref*{UE_FLOPs} illustrates the training parameters and FLOPs for different modules at each UE side. Bar plots represent the parameters and FLOPs of every module, while line plots show the total parameters and FLOPs for every vendor. In the JEFPNet architecture, the UE side consists only of a joint CE and CSI compression module, denoted as ``Joint CE and CSI Compression''. In contrast, the partially separated SEFPNet and fully separated traditional architectures with MMSE and WF algorithms, denoted as MMSE, implement CE and CSI compression as distinct modules, denoted as ``CE'' and ``CSI Compression'', respectively. Statistical results reveal that the CE module in the separate architectures requires significantly more parameters and FLOPs. Additionally, the CSI compression module for feature extraction and CSI compression further increases the computational demand and parameter storage overhead. By comparison, the proposed JEFPNet architecture requires only 2.63\% of the parameters and 0.019\% of the FLOPs compared to the separate architectures, making it more suitable for hardware-constrained UEs. 

Fig~\subref*{BS_FLOPs} displays the parameters and FLOPs at BS with $K=2$. Since JEFPNet and SEFPNet share the same network architecture at the BS, their parameters and FLOPs, labeled as ``Joint CSI Reconstruction and Precoding Design'', are identical. For the joint CSI reconstruction and precoding design module, since its network architecture depends on the number of users, both the parameters and FLOPs shown in Fig.~\subref*{BS_FLOPs} correspond to the case of two users. For the traditional architecture, CSI reconstruction at the BS is implemented using $K$ decoders with shared parameters, labeled as ``CSI Reconstruction''. Therefore, only one decoder’s parameters need to be stored and counted for statistics, which is shown on the left of Fig.~\subref*{BS_FLOPs}. In the separate architecture, we employ MMSE for precoding design and the WF algorithm for power allocation, collectively referred to as ``MMSE''. Since the MMSE and WF algorithms do not include trainable parameters, their parameter count is 0. For the FLOPs statistics shown on the right side of Fig.~\subref*{BS_FLOPs}, the bar plot represents the FLOPs of every modular in the MMSE architecture, while the line plot corresponds to the total FLOPs at the BS vendor. As shown in Fig.~\subref*{BS_FLOPs}, {to avoid the complexity of model monitoring and switching at the BS, the proposed JEFPNet introduces a user-number-adaptive design. Consequently, in the two-user case, JEFPNet requires more training parameters and incurs higher FLOPs on the BS side due to the larger $K_{\max}$. However, without user-number adaptability design, when $K_{\max}=6$, the BS must store multiple models corresponding to different user numbers. Specifically, for $K=2$ to $6$, the non-adaptive network requires five separate models with a total parameter count of 1.2912M, whereas the adaptive design reduces this requirement to only 20\% of that amount. Moreover, the non-adaptive network also necessitates training different encoder networks, leading to additional model delivery or storage overheads for the UE side.}
% the joint CSI reconstruction and precoding design in JEFPNet requires more training parameters and FLOPs than the separate architecture. The increase in storage overhead is not an issue for the BS equipped with sufficient hardware capabilities. Comparatively, the reduction in FLOPs decreases CSI feedback and precoding design latency, thereby enhancing precoding reliability and mitigating the effects of CSI aging.
\begin{figure}[!t]
	\centering
	\includegraphics[width=3.4in]{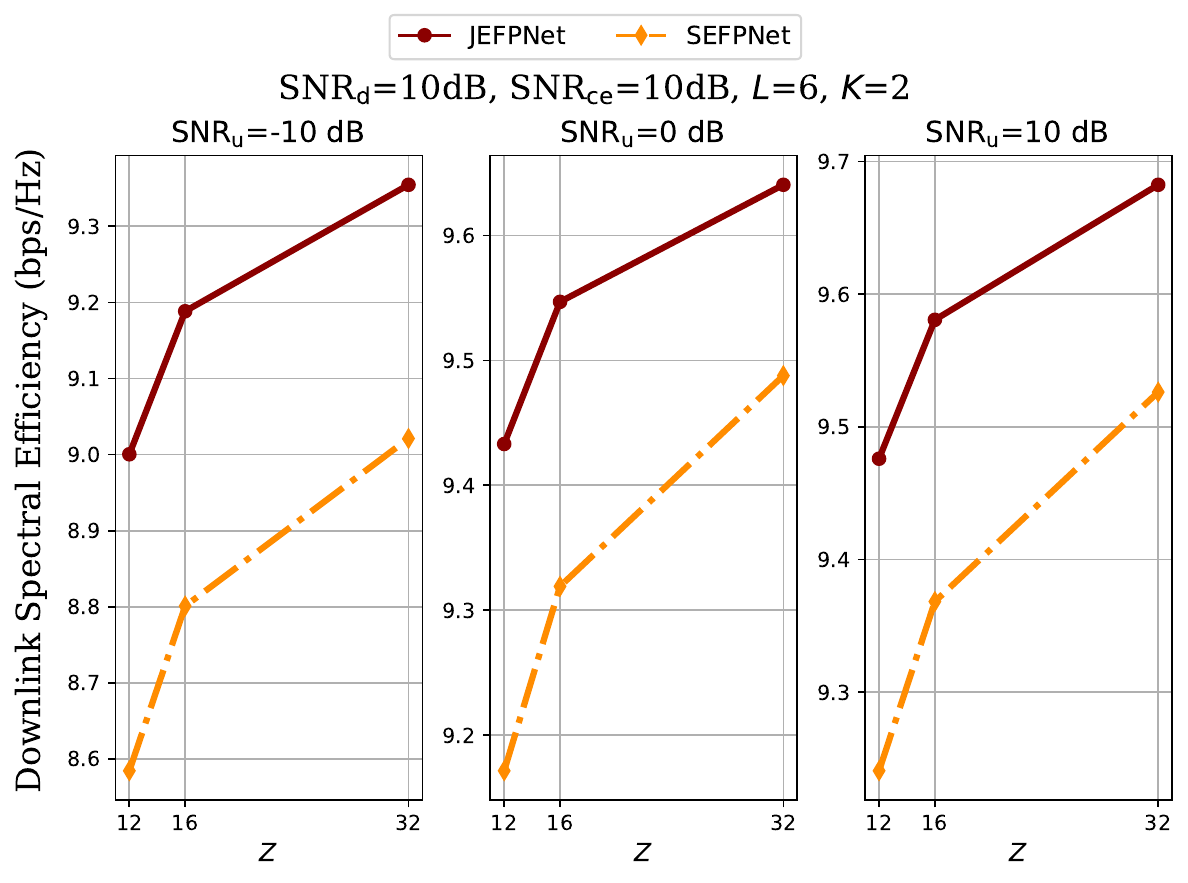}
	\caption{Testing the downlink spectral efficiency of joint CE-feedback-precoding networks under partially separate and joint architectures as affected by the amount of feedback.}
	\label{7_Diff_ed}
\end{figure}
\subsection{Scalability Tests}
In this subsection, we firstly experimentally verify the scalability of the proposed network under varying feedback overheads. First, we fix the $\mathrm{SNR_{ce}}$ of the CE phase to 10 dB, and the CE overhead ($L$) to 6. The maximum number of user $K_{\max}$ set to 6, and the evaluated number of user $K$ is 2. We compare the downlink spectral efficiencies of different network architectures with single-user CSI feedback overhead ($Z$) values of 12, 16, and 32. The experimental results, shown in Fig.~\ref{7_Diff_ed}, indicate that as feedback overhead increases, the downlink spectral efficiency for both joint and separate architectures gradually improves. {As the feedback bandwidth increases, the performance gap between the two methods gradually narrows.
Specifically, the downlink spectral efficiency of JEFPNet exceeds that of SEFPNet by 0.416 bps/Hz when $Z = 12$, 0.387 bps/Hz when $Z = 16$, and 0.333 bps/Hz when $Z = 32$, under $\mathrm{SNR_{u}}=-10$ dB.}
% Under low feedback bandwidth $Z$ and $\mathrm{SNR_u}$ conditions, the proposed JEFPNet achieves superior performance. At $Z = 12$ and $\mathrm{SNR_u} = -10$~dB, JEFPNet outperforms JFPNet by 0.31~bps/Hz, owing to its joint CE and CSI feedback design. In limited feedback bandwidth scenarios, extracting features directly from the received pilot signals yields more effective task-oriented information than extracting features from the full CSI. As the feedback bandwidth and channel quality improve, this advantage gradually diminishes but remains superior to that of SEFPNet, which also accounts for CE errors.}

\begin{figure}[!t]
	\centering
	\includegraphics[width=3.4in]{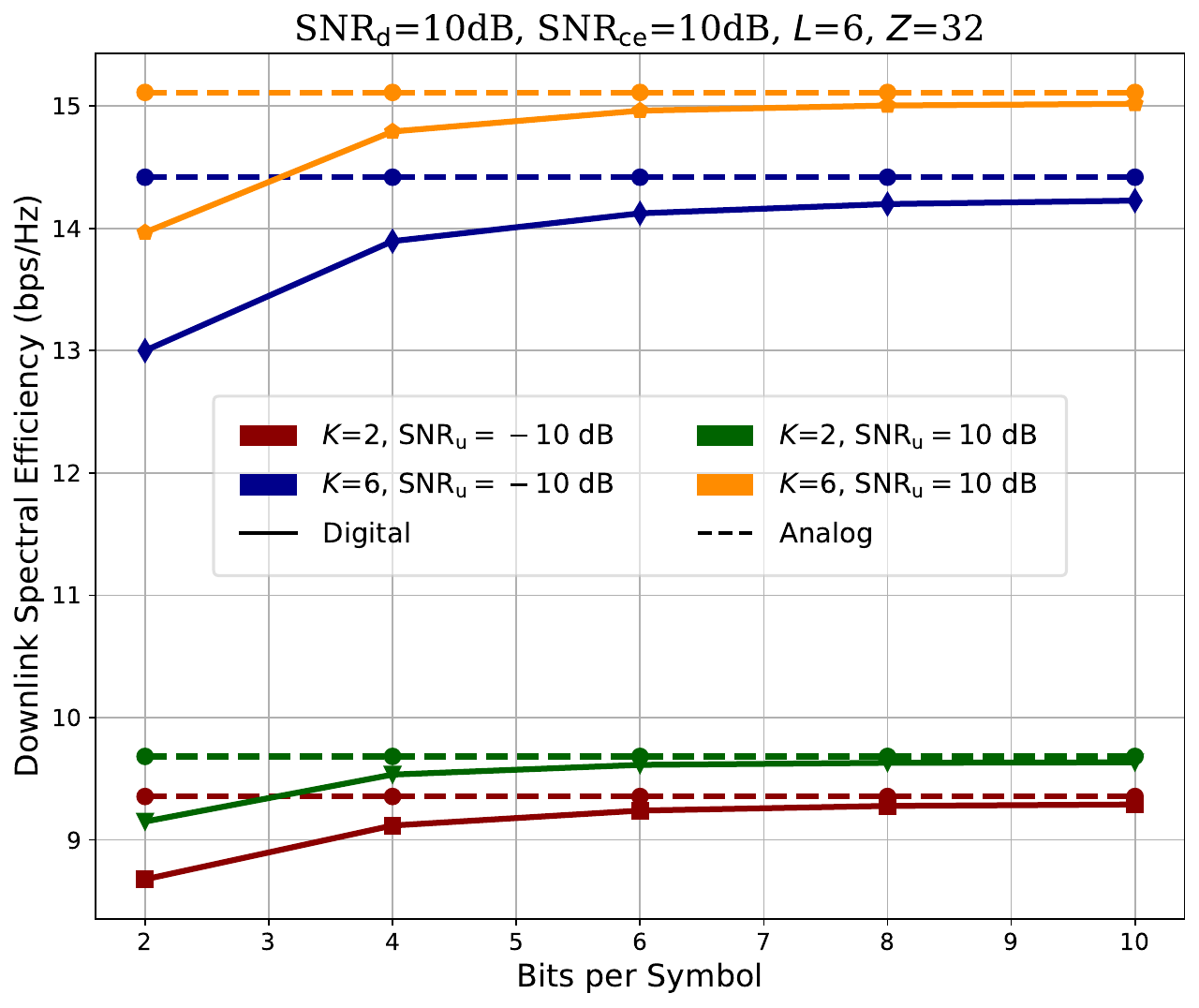}
	\caption{CSI rate-efficiency bounds.}
	\label{8_Digital_vs_Analog}
\end{figure}

{In this paper, we adopt the DJSCC network architecture in the analog domain. To facilitate deployment in existing communication systems and maintain compatibility with current hardware, we assess the rate–efficiency bound of the proposed network in a digital communication system, analogous to the CSI rate–distortion bounds presented in \cite{rate_distortion_1,rate_distortion_2}. Specifically, the downlink CSI features compressed at the UE side are mapped to the nearest power-normalized quadrature amplitude modulation (QAM) constellation points for transmission, while the BS utilizes the received constellation points for multiuser precoding design. In this part of the simulation experiments, we consider a CE phase with $\mathrm{SNR_{ce}}=10$ dB, a pilot length $L=6$, and modulation schemes including 4-QAM, 16-QAM, 64-QAM, 256-QAM, and 1024-QAM. These correspond to symbol quantization levels of 2, 4, 6, 8, and 10 bits, respectively. In this way, the feedback bandwidth $Z$ is maintained at a constant value of 32. The corresponding simulation results are presented in Fig.~\ref{8_Digital_vs_Analog}.}

{The simulation results in Fig.~\ref{8_Digital_vs_Analog} demonstrate that the performance of JEFPNet in the digital domain gradually converges to that in the analog domain as the number of quantization bits increases, regardless of the number of users or uplink feedback $\mathrm{SNR_u}$. For instance, under the condition $K=2$, $\mathrm{SNR_{u}}=10$~dB, and 1024-QAM, the performance loss is only 0.048~bps/Hz. Similarly, with $K=6$ under the same conditions, the loss is limited to 0.093~bps/Hz. This improvement arises because increasing the number of quantization bits leads to a higher modulation order, resulting in a denser constellation and thereby reducing the loss and error introduced by mapping the analog signal to the constellation points.}

\subsection{Scenario Shift Evaluation}

\begin{table}[!t]
\setlength\tabcolsep{4.5pt}
    \caption{Scenario shift testing in UMi and UMa environments with downlink spectral efficiency.}
    \renewcommand{\arraystretch}{1.15}
    \centering
    \begin{tabular}{c|c|cc|cc|cc}
    \toprule
                          bps/Hz& $\mathrm{SNR_u}$ & \multicolumn{2}{c|}{$\mathrm{-10}$~dB}    & \multicolumn{2}{c|}{$\mathrm{0}$~dB}      & \multicolumn{2}{c}{$\mathrm{10}$~dB}     \\ \midrule 
    $K$ &  
\diagbox{\textbf{Test}}{\textbf{Train}}& \multicolumn{1}{c}{\textbf{UMa}} & \textbf{UMi} & \multicolumn{1}{c}{\textbf{UMa}} & \textbf{UMi} & \multicolumn{1}{c}{\textbf{UMa}} & \textbf{UMi} \\ \midrule
    \multirow{2}{*}{2} & \textbf{UMa}  & \multicolumn{1}{c}{9.41}  & 9.36  & \multicolumn{1}{c}{9.78}  & 9.74  & \multicolumn{1}{c}{9.81}  & 9.78  \\  
                         & \textbf{UMi}  & \multicolumn{1}{c}{9.57}  & 9.47  & \multicolumn{1}{c}{10.33} & 10.18 & \multicolumn{1}{c}{10.41} & 10.26 \\ \midrule
    \multirow{2}{*}{4} & \textbf{UMa}  & \multicolumn{1}{c}{12.52} & 12.36 & \multicolumn{1}{c}{13.12} & 12.96 & \multicolumn{1}{c}{13.18} & 13.02 \\ 
                         & \textbf{UMi}  & \multicolumn{1}{c}{12.57} & 12.48 & \multicolumn{1}{c}{13.89} & 13.76 & \multicolumn{1}{c}{14.04} & 13.91 \\ \toprule
    \end{tabular}
    \label{Shift_Performance}
    \end{table}

{Investigating the scenario-shift problem faced by AI models is critical for next-generation communication systems that integrate AI with communication technologies, and it has been identified as a key research challenge in the 6G standardization process by 3GPP~\cite{3gpp38843}. Accordingly, in this set of experiments, we evaluate the impact of scenario shift on the performance of the proposed network. Specifically, we introduce additional Urban Micro (UMi) scenarios in comparison to the previously considered UMa scenarios. The UMi dataset is also generated using QuaDriGa following 3GPP TR~38.901 and TR~38.873. Except for scenario-specific parameters, such as the BS height, all other settings remain consistent with those of the UMa dataset to ensure fairness in comparison. We train JEFPNet under both UMa and UMi scenarios and subsequently test the trained models in both environments to assess the impact of scenario shifts on network performance, using downlink spectral efficiency as the evaluation metric. In this part of the experiment, we set $K_{\mathrm{max}}=4$, $Z=32$, $L=6$, and ${\mathrm{SNR_{ce}}}=10$~dB. We then evaluated the impact of scenario shift on JEFPNet performance in both UMi and UMa scenarios, considering $\mathrm{SNR_{u}}\in\{-10~\text{dB}, 0~\text{dB}, 10~\text{dB}\}$ and $K\in\{2,4\}$. The experimental results are presented in Table.~\ref{Shift_Performance}.}

{The experimental results indicate that, without scenario shifts, performance in the UMi scenario is better than in the UMa scenario. For example, with $K=2$ and $\mathrm{SNR_{u}}=-10$~dB, JEFPNet in the UMi scenario outperforms that in the UMa scenario by 0.16~bps/Hz. Similarly, with $K=6$ and $\mathrm{SNR_{u}}=10$~dB, the performance improvement reaches 0.86~bps/Hz. This advantage arises because the spatial characteristics of the UMi scenario are more favorable than those of the UMa scenario, enabling convergence to a better solution under identical training conditions. A similar phenomenon was observed in \cite{UMi_UMa}, where the CE error of the UMi scenario is smaller than that of the UMa scenario under the same conditions. In the presence of scenario shifts, the network trained in the UMa scenario performs worse than those trained in the UMi scenario when evaluated in the UMi scenario, although the performance gap narrows at lower $\mathrm{SNR_{u}}$. For instance, at $K=2$ and $\mathrm{SNR_{u}}=-10$~dB, the gap is only 0.16~bps/Hz. Conversely, networks trained in the UMi scenario often achieve superior performance than those trained in the UMa scenarios when tested in the UMa scenario, despite a slight decline, owing to their more favorable convergence positions.}
\section{Conclusion}
\label{section4}
In this paper, considering the channel coding and feedback channel, we propose an end-to-end architecture for CSI acquisition and precoding design in MU MIMO-OFDM systems with user-number adaptability, called JEFPNet. At the UE side, a joint CE and compression network is proposed to compress and extract only the semantic information relevant to precoding from received pilot signals, rather than the entire CSI. This approach eliminates the overhead of explicit CE and minimize information loss. At the BS, a proposed joint MU CSI reconstruction and Transformer-based precoding design network directly designs the precoding vectors from the received semantic information fed from each user. By leveraging a multi-head attention mechanism, the precoding design module can be adapted to varying numbers of users without substantially increasing the BS parameter count, thereby reducing model storage requirements, distribution overhead, and switching latency.
Additionally, a pilot learning matrix is employed at the BS to enhance the extraction efficiency of semantic information and maximize its utilization. The entire network is trained end-to-end with the objective of maximizing the downlink spectral efficiency. Multiple modules are jointly optimized to ensure that only semantic information related to the optimization objective is extracted and transmitted throughout the transmission process. 

Experimental results demonstrate that the proposed network achieves a higher downlink spectral efficiency under the same spectrum resource overhead. Furthermore, JEFPNet significantly reduces both storage overhead and computational complexity at the UE side, which is more suitable for UE side without powerful storage capacity and computational ability. Finally, we evaluate the scalability of the proposed network under varying feedback overheads and its adaptability to different numbers of users, demonstrating that the performance of the network under the proposed joint architecture consistently outperforms that of the separate architecture.

\normalem
\bibliographystyle{IEEEtran}
\bibliography{myref}

% \newpage

 

% \end{IEEEbiography}



% \vfill

\end{document}